\documentclass[iop]{emulateapj}
\slugcomment{{\sc Accepted to ApJ:} November 7, 2011}
\usepackage{psfrag}
\newcommand\nion[2]{#1\,\lowercase{{\sc #2}}}
\newcommand\wave[1]{\mbox{$\lambda$#1\,\AA}}
\def\kmsec{\mbox{km~s$^{\rm -1}$}}
\def\teff{\mbox{T$_{\rm eff}$}}
\def\BmV0{\mbox{$(B-V)^{\rm 0}$}}
\def\VmK0{\mbox{$(V-K)^{\rm 0}$}}
\def\MV0{\mbox{$M_{\rm V}^{\rm 0}$}}

\def\msun{M$_{\odot}$}

\def\ciso{$^{12}$C/$^{13}$C}
\def\lum{\log(L/L_{\odot})}
\def\teff{T_{\rm eff}}
\def\tini{T_{\rm ini}}

\def\logg{\log~g}
\def\loggi{\log~g_{\rm ini}}

\def\fehi{\rm[Fe/H]_{ini}}
\def\feh{\rm[Fe/H]}

\def\cfe{\rm[C/Fe]}
\def\nfe{\rm[N/Fe]}
\def\ofe{\rm[O/Fe]}

\def\simgt{\lower.5ex\hbox{$\; \buildrel > \over \sim \;$}}
\def\simlt{\lower.5ex\hbox{$\; \buildrel < \over \sim \;$}}
\def\ali{A({\rm Li})}
\def\vsini{v{\rm sin}i}

\shorttitle{Metal-Poor Li-rich Giants in RAVE}
\shortauthors{Ruchti et al.}

\begin{document}

\title{Metal-Poor Lithium-Rich Giants in the RAVE Survey\altaffilmark{1}}

\author{
Gregory~R.~Ruchti\altaffilmark{1,2}
Jon~P.~Fulbright\altaffilmark{1},
Rosemary~F.~G.~Wyse\altaffilmark{1},
Gerard~F.~Gilmore\altaffilmark{3,4},
Eva~K.~Grebel\altaffilmark{5},
Olivier~Bienaym\'{e}\altaffilmark{6},
Joss~Bland-Hawthorn\altaffilmark{7},
Ken~C.~Freeman\altaffilmark{8},
Brad~K.~Gibson\altaffilmark{9,10},
Ulisse~Munari\altaffilmark{11},
Julio~F.~Navarro\altaffilmark{12},
Quentin~A.~Parker\altaffilmark{13,14,15},
Warren~Reid\altaffilmark{14},
George~M.~Seabroke\altaffilmark{16},
Arnaud~Siebert\altaffilmark{6},
Alessandro~Siviero\altaffilmark{17,18},
Matthias~Steinmetz\altaffilmark{18},
Fred~G.~Watson\altaffilmark{13},
Mary~Williams\altaffilmark{18},
Tomaz~Zwitter\altaffilmark{19,20}
}

\affiliation{
$^1$Bloomberg Center for Physics \& Astronomy, Johns Hopkins University, 3400 North Charles Street, Baltimore, MD 21218, USA: gruchti@mpa-garching.mpg.de\\
$^2$Current Address: Max Planck Institut f\"ur Astrophysik, Postfach 1317, Karl-Schwarzschild-Str. 1, D-85748 Garching, Germany\\
$^3$Institute of Astronomy, University of Cambridge, Madingley Road, Cambridge CB3 0HA, UK\\
$^4$ Astronomy Department, Faculty of Science, King Abdulaziz University, P.O.Box 80203, Jeddah 21589, Saudi Arabi\\
$^5$Astronomisches Rechen-Institut, Zentrum f\"ur Astronomie der Universit\"at Heidelberg, M\"onchhofstr.\ 12--14, 69120 Heidelberg, Germany\\
$^6$Observatoire de Strasbourg, 11 Rue de L'Universit\'{e}, 67000 Strasbourg, France\\
$^7$Sydney Institute for Astronomy, School of Physics A28, University of Sydney, NSW 2006, Australia\\
$^8$RSAA Australian National University, Mount Stromlo Observatory, Cotter Road, Weston Creek, Canberra, ACT 2611, Australia\\
$^9$Jeremiah Horrocks Institute for Astrophysics \& Super-computing, University of Central Lancashire, Preston, PR1 2HE, UK\\
$^{10}$Department of Astronomy \& Physics, Saint Marys University, Halifax, B3H 3C3, Canada\\
$^{11}$INAF Osservatorio Astronomico di Padova, Via dell'Osservatorio 8, Asiago I-36012, Italy\\
$^{12}$Department of Physics and Astronomy, University of Victoria, P.O. Box 3055, Station CSC, Victoria, BC V8W 3P6, Canada\\
$^{13}$Australian Astronomical Observatory, Coonabarabran, NSW 2357, Australia\\
$^{14}$Department of Physics and Astronomy, Macquarie University, Sydney, NSW 2109, Australia\\
$^{15}$Macquarie Research Centre for Astronomy, Astrophysics \& Astrophotonics, Macquarie University, Sydney, NSW 2109, Australia\\
$^{16}$Mullard Space Science Laboratory, University College London, Holmbury St. Mary, Dorking RH5 6NT, UK\\
$^{17}$Department of Astronomy, Padova University, Vicolo dell'Osservatorio 2, Padova 35122, Italy\\
$^{18}$Leibniz-Institut f\"ur Astrophysik Potsdam (AIP), An der Sternwarte 16, 14482 Potsdam, Germany\\
$^{19}$Faculty of Mathematics and Physics, University of Ljubljana, Jadranska 19, 1000 Ljubljana, Slovenia\\
$^{20}$Center of Excellence SPACE-SI, Askerceva cesta 12, 1000 Ljubljana, Slovenia
}
\altaffiltext{1}{Based on observations taken at the Keck, Apache Point, Las Campanas, and La Silla (ESO proposal ID: 082.B-0484) Observatories.}

\begin{abstract}
We report the discovery of eight lithium-rich field
giants found in a high resolution spectroscopic sample of over 700
metal-poor stars ($\feh<-0.5$) selected from the RAVE survey.  The
majority of the Li-rich giants in our sample are very metal-poor ($\feh\simlt-1.9$), and
have a Li abundance (in the form of $^7$Li), $\ali=\log(n({\rm Li})/n({\rm H}))+12$, between 2.30 and 3.63, well above the typical upper
red giant branch limit, $\ali<0.5$, while two stars, with
$\ali\sim1.7-1.8$, show similar lithium abundances to normal giants at the same
gravity. We further included two metal-poor, Li-rich globular cluster
giants in our sample, namely the previously discovered
M3-IV101 and newly discovered (in this work) M68-A96.
This comprises the largest sample of metal-poor Li-rich
giants to date.  We performed a detailed abundance analysis of all
stars, finding that the majority our sample stars have elemental abundances similar to that of Li-normal halo giants.  Although the evolutionary
phase of each Li-rich giant cannot be definitively determined, the Li-rich
phase is likely connected to extra mixing at the red giant branch bump or early asymptotic giant branch that triggers cool bottom processing in which the bottom of the outer convective envelope is connected to the H-burning shell in the star.  The surface of a star becomes Li-enhanced as $^7$Be (which burns to $^7$Li) is transported to the stellar surface via the Cameron-Fowler mechanism.  We discuss and discriminate among several models for the extra mixing that can cause Li-production, given the detailed abundances of the Li-rich giants in our sample. 
\end{abstract}


\section{Introduction}
\label{sec-intro}

Lithium (Li) plays a special role in our understanding of the
Universe.  Li, in the form of $^7{\rm Li}$, is one of four isotopes
synthesized immediately after the big bang \citep{steigman07}.  The latest estimate
of the cosmic baryon density from WMAP,
$\Omega_bh^2=0.02273\pm0.00062$ \citep{dunkley09}, implies a
primordial abundance\footnote{$\ali=\log\frac{n({\rm Li})}{n({\rm H})}+12$} of $\ali=2.72\pm0.06$, using the updated reaction
rates for $^3{\rm He}(\alpha,\gamma)^7{\rm Li}$ in the standard big
bang nucleosynthesis calculations \citep{cyburt08}.  This value
of the primordial Lithium-7 abundance is significantly higher than
that derived for metal-poor stars, for which $\ali=2.0-2.4$ for
$\feh=-3.5$ to -1.0~dex
\citep{spite82a,spite82b,ryan01,melendez04,charbonnel05,asplund06,bonifacio07,aoki09,hosford09,sbordone10}.
Lithium is expected to be destroyed in stars, creating helium, in
regions where the temperature exceeds a few times $10^6$~K. However,
the large amplitude of the discrepancy with the predicted primordial value
of Lithium-7, together with its apparent constancy over a range of stellar effective temperatures and gravities (the `Spite Plateau'),  has stimulated much interest into both Lithium destruction and production in 
stellar interiors and into extensions of the Standard Model of particle physics.

The abundance of Li in stellar atmospheres is a very useful probe
of the structure of the stellar interior and the physical processes
taking place there. The fragile Li nucleus is readily
destroyed when the material is exposed to temperatures exceeding
$2.6\mathsf{x}10^6$~K, so that strong Li depletion is usually observed
in any star whose surface convection zone extends deep enough.  As soon as
a star moves beyond the sub-giant branch, convective depletion brings
$\ali$ down by more than one order of magnitude \citep{p93,g00}, and
$\ali<0.5$ is typical for  stars on the upper red giant branch (RGB) \citep{lind09b}.  Any giant with a Li abundance above this value is
considered Li-rich.  About 1\% of solar-metallicity giant stars show
large Li abundances \citep[][]{brown89}.  These giants pose
a serious problem for standard stellar evolution models, and have triggered a widespread interest in Li-production in giant stars.

Standard models of stellar evolution predict that Li can be produced in the interior, but it is immediately destroyed by nuclear burning, as explained above.  Further, pre-existing Li in the surface layers of a star is burned away due to dilution.  However, should there exist efficient, extra mixing between the surface and the Li-forming layers, $^7$Li (or its progenitor $^7$Be) can be brought to the cool layers before it burns.  Many have tried to explain these phenomena using both internal and external processes.  

\citet{cameron55} and \citet{cameron71} first proposed a  
mechanism in which Li could be produced (and survive) by $^7$Be-transport to the surface of intermediate-mass asymptotic giant branch (AGB) stars, known as the Cameron-Fowler mechanism.  At this stage the outer convective envelope is in contact with the H-burning shell where $^3$He-enrichment has taken place from the proton-proton reaction chain.  The $^3$He is transported to regions with temperatures high enough to burn it to $^7$Be by the ${\rm ^3He(\alpha,\gamma)^7Be}$ reaction.  The $^7$Be is then swept up to the stellar surface where it decays to $^7$Li by electron captures (${\rm ^7Be({\it e}^-,\nu)^7Li}$).  This process of $^3$He transport from the inner H-burning shell and subsequent burning to $^7$Be is otherwise known as ``hot bottom burning" \citep[cf.][and references therein]{forestini97}.

Later, \citet{sackmann99} showed that the Cameron-Fowler mechanism can also occur for low-mass giants evolving on the RGB due to extra deep mixing and ``cool bottom processing" (CBP), in which material from the cool-bottom of the outer convective envelope reaches temperatures in which the $^3$He is burned.  This process was first postulated to explain the abundances of $^{13}$C in AGB and RGB stars \citep{boothroyd95}.  To produce enhanced $^7$Li on the surface, CBP requires high mixing rates ($\dot{M}_p\simgt10^{-7}~M_{\odot}\rm{yr}^{-1}$).  Depending on this mixing rate, a star can achieve a Li-enhancement upwards of $\ali\sim4$.  \citet{sackmann99} suggest that the rarity of Li-rich giants implies that few stars can achieve high enough mixing rates to drive Li-production or that the episode of rapid mixing is brief.  \citet{boothroyd99} also note that low-mass, metal-poor RGB stars should undergo more aggressive CBP than metal-rich giants, since the extra mixing will reach higher temperatures.  Further, if the $^3$He is not fully depleted during the evolution up the RGB, then CBP can also occur on the AGB.  In low-mass AGB stars, the outer envelope and H-burning shell are not in contact.  Deep extra-mixing via CBP can connect the two regions, allowing the Cameron-Fowler mechanism to occur in low-mass AGB stars \citep{nollett03}.

More recently, thermohaline mixing \citep{charbonnel05,charbonnel07} and magneto-thermohaline mixing \citep{denissenkov09} have been proposed as sources of extra mixing to drive Li-production on the surface-layers of stars.  On the other
hand, extra mixing processes at the RGB-bump \citep{c95} may also induce a
so-called ``Li-flash" \citep{p01}.  In this case, $^7$Be is transported to the surface layers of a star, while the star is also rapidly increasing in luminosity.  This model was later challenged by \citet{dh04}, who independently found that canonical extra mixing in stars cannot produce a Li-flash.  In their more recent investigations, \citet{palacios06} were still not able to validate the Li-flash model.  

Another scenario invokes mass-loss
mechanisms on the RGB, which are accompanied by extra mixing that increases the
Li-abundance \citep{dla96,dla00}.  Note, however, that subsequent searches \citep[e.g.,][]{fekel98,jasniewicz99} did not detect any Li-rich stars in samples of giants with far-infrared excess.  Finally, external angular momentum
from a companion object (brown dwarf or giant planet) may induce extra
mixing needed to drive the Cameron-Fowler mechanism \citep{dh04}, which in turn leads to increased Li.

These models are described in detail in \S\ref{sec-lisy}, but common features
among these theories is that the Li-enrichment on the surface layers of a star can reach values $\sim2$~dex higher than the Spite Plateau.  Further, the Li-enrichment phase is short, lasting about 2~Myr \citep[see \S\ref{sec-lisy} and e.g.,][]{dh04}.  Another important aspect of these models is that material from the CNO burning regions is also transported to the stellar surface.  The distinguishing features among the models
include the mixing mechanism and the timing of the mixing/burning
episodes along a star's evolution, which can affect the amount of CNO-material brought to the stellar surface.  These could also have an effect on the abundances of
heavier elements, such as r-, s-, and possibly p-process elements
present in the star's atmosphere.  Detailed elemental abundances of
Li-rich giants will therefore provide insight into mixing and
nucleosynthesis processes within evolved stars, and will ultimately
further our understanding of the origins of these peculiar stars.

Li-rich giants have been discovered in the field \citep{cb00,roed,kumar09,kumar11,monaco11} and Galactic bulge \citep{uttenthaler07,gonzalez09}, as well as in globular clusters \citep{cfg98,kraft99,ssk99} and dwarf spheroidal galaxies \citep{dominguez04,monaco08}.  These samples contain stars on both the RGB and AGB that range in mass ($\sim0.8-5~M_{\odot}$) and have metallicities from solar down to $\sim-2$~dex.  It is important to note, however, that the vast majority of known Li-rich giants have metallicities near solar, while few ($\sim4-5$) Li-rich giants have been discovered with $\feh<-1$.

In their compilation of near solar-metallicity stars, \citet{cb00} found that Li-rich giants primarily cluster around two regions in the
color-magnitude diagram.  They associated low-mass Li-rich giants with the RGB luminosity bump (RGB-bump), while those with intermediate masses were assumed be evolving on the early AGB.  This is convenient, because in both regions extra mixing can be triggered after the molecular weight discontinuity from the first dredge-up is erased (see above).  \citet{cb00} suggested that the localization of the Li-rich giants into two groups argued against the interaction with a companion object as the cause for Li-production.  This two-region picture, however, has been challenged by the discovery of low-mass Li-rich giants that lie near the RGB-tip \citep[e.g.][]{kraft99,monaco08}, where \citet{cb00} had classified solar-metallicity stars as intermediate-mass AGB.  Further, in a study of Li-rich giants in the thick disk of the Milky Way, \citet{monaco11} found that the Li-rich giants in their sample did not fall in either group defined by \citet{cb00}.

Metal-poor, low-mass Li-rich giants near the RGB-tip further confuse the situation.  It is possible that these stars have not yet reached the AGB.  Indeed,
\citet{kraft99} found that the Li-rich giant, M3-IV101, had a luminosity placing it near the RGB-tip, but was most likely an RGB star according to the color-magnitude diagram of M3, using high-precision photometry.  How can we distinguish between the RGB and AGB for metal-poor stars?  A possible discriminant (in addition to the CNO abundance differences between different theories of Li-production) is that AGB stars that have begun the third dredge-up typically show enhancements in the s-process elements as
compared to RGB stars.  As stated above, the number of known metal-poor Li-rich giants is quite small.  A clear classification of the evolutionary
phase of these stars is absolutely critical for understanding the
processes that create the Li.  It is therefore important to identify
and analyze more metal-poor Li-rich giants.

We have discovered nine candidate Li-rich metal-poor giants.  Eight of the stars
were part of high-resolution observations of metal-poor stars selected from the 
Radial Velocity Experiment survey \citep[RAVE,][]{rave}, and one, in the very 
metal-poor globular cluster M68, was found by us independently of RAVE. In this paper, we report on the abundance properties of these stars and investigate possible signatures for each star's evolutionary stage, and look for supporting evidence of the mechanism for enhanced Li-production.

\section{Observations}

The Li-rich stars reported here were among over 700 candidate metal-poor stars selected for high-resolution observations \citep{ruchti10,f10},
based on data obtained by the RAVE survey,  with
the exception of M68-A96, whose Li-rich nature was discovered during
observations of stars in that globular cluster.  The full details of
the high-resolution observations and reductions of the RAVE stars can
be found in the papers cited above, but some information is given in Table~\ref{tab-obs}.  M68-A96 was observed with the
echelle spectrograph on the Ir{\'e}n{\'e}e du Pont 2.5-m telescope at
the Las Campanas Observatory.  For comparison purposes, we also
analyzed a blue Keck/HIRES spectrum taken in March 1999 of the
previously-known Li-rich giant M3 IV-101.  Note that this spectrum is not the same as that used by \citet{kraft99}.

The spectrographs used for our observations deliver a resolving power greater than 30,000.  The 
S/N level of the observed spectra are quite good:  nearly all have S/N ratios
greater than 100 per pixel. 
With the exception of the UCLES spectrum, the wavelength coverage goes 
from below 4000 \AA{} to beyond  8000~\AA{} (for UCLES the range is roughly
4460--7260 \AA), although there are some gaps in coverage.  In each case the 
data were reduced using standard reduction methods for echelle data, utilizing
pipeline reduction programs when available.

During routine inspection of the $\sim700$ spectra taken, we noticed 
that the 
6708~\AA{} \nion{Li}{I} lines in some of the stars' spectra were unusually strong. 
For example, the  \nion{Li}{I} line
in the star J142546.2-154629 has an equivalent width (EW) of 
$\sim$~540~m\AA{} (see \S\ref{sec-li}).  This is roughly twice as strong as each of the Na D lines 
in this star.  In several of the other spectra, the 6708~\AA{} line 
appeared on two adjacent orders, so it was very unlikely the feature was an artifact introduced by some feature of the observation or reductions. The 6103~\AA{} \nion{Li}{I} line was also visible in most of these anomalous stars, confirming the high Li abundances.

\begin{center}
\setlength{\tabcolsep}{0.04in}
\begin{deluxetable*}{rrrrcrrr}
\tablecolumns{8}
\tabletypesize{\scriptsize}
\tablewidth{0pc}
\tablecaption{Observational Data}
\tablehead{
\colhead{Star} & \colhead{RA\tablenotemark{\it a}} & \colhead{DEC\tablenotemark{\it a}} & \colhead{Mag.\tablenotemark{\it b}} & \colhead{Obsdate} & \colhead{Observatory\tablenotemark{\it c}} & \colhead{Instrument} & \colhead{S/N\tablenotemark{\it d}} \\
& \colhead{($^{\circ}$)} & \colhead{($^{\circ}$)} & & \colhead{(yyyymmdd)} & & & \colhead{(${\rm pix}^{-1}$)}
}
\startdata
C1012254-203007 &  153.106 & -20.502 & 11.6 & 20070506 & LCO & MIKE & 100 \\
J043154.1-063210 &    67.976 &   -6.536 & 10.4 & 20071219 & APO & ARCES & 100 \\
J142546.2-154629 & 216.443 & -15.775 & 9.8 & 20090212 & La Silla & FEROS & 130 \\
J195244.9-600813 & 298.187 & -60.137 & 10.1 & 20081016 & LCO & MIKE & 160 \\
T5496-00376-1 & 155.025 & -13.095 & 9.1 & 20070421 & APO & ARCES & 130 \\
T6953-00510-1 & 327.969 & -22.935 & 9.9 & 20081016 & LCO & MIKE & 140 \\
T8448-00121-1 & 348.250 & -45.119 & 10.0 & 20070922 & AAT & UCLES & 100 \\
T9112-00430-1 &  312.012 & -65.966 & 10.4 & 20081016 & LCO & MIKE & 140 \\
M3-IV101 & 196.259 & 28.303 & (13.2) &  19990323 & Keck & HIRES & 80 \\
M68-A96 & 180.659 & -26.717 & (13.0) &  20040106 & LCO & du Pont Echelle & 100 
\enddata
\label{tab-obs}
\tablenotetext{a}{equinox 2000}
\tablenotetext{b}{The $I$-magnitudes from the RAVE database are given here for the eight RAVE stars.  $V$-magnitudes are given in parentheses for M3-IV101 \citep{johnson56} and M68-A96 \citep{alcaino77}.}
\tablenotetext{c}{LCO=Los Campanas Observatory, APO=Apache Point Observatory, AAT=Australian Astronomical Telescope}
\tablenotetext{d}{Estimated between \wave{5500-6000}.}
\end{deluxetable*}
\end{center}

\section{Analysis}

\subsection{Stellar Parameters}

The abundance determinations were achieved with the MOOG analysis program \citep{moog}, using 1-D, plane-parallel Kurucz
model atmospheres\footnote{See http://kurucz.harvard.edu/.} under the assumption of static equilibrium and LTE.  Stellar parameters were derived following a variation of the methods described in \citet{ruchti10,ruchti11}.  The initial effective temperature, $\tini$, was set by using the excitation temperature method based on \nion{Fe}{I} lines.  The initial value of the surface gravity ($\loggi$) was set using the ionization equilibrium criterium utilizing the iron abundance derived by both \nion{Fe}{I} and \nion{Fe}{II} lines.  The initial $\feh$ value of the stellar atmosphere for each star was chosen to match the [\nion{Fe}{II}/H] value derived from the analysis.  The value of the microturbulent velocity ($v_t$) was set to minimize the magnitude of the slope of the relationship between the iron abundance derived from \nion{Fe}{I} lines and the value of the reduced width of the line.

In \citet{ruchti10,ruchti11}, we found during our analysis of the spectra for several globular cluster giants and giant
stars selected from the \citet{fulbright00} sample that
 the effective temperature
estimate from the excitation method showed an offset, that correlated
with $\fehi$, when compared to photometric temperature estimates
\citep[using the 2MASS color-temperature transformations
of][]{ghernandez09}.  We have since found similar results  using the
color-temperature transformations of Casagrande et al. (2011, private
communication).   We therefore applied the
temperature corrections described in \citet{ruchti11} to the Li-rich
candidates in our current sample.  The analysis described above was
then performed again, but with the effective temperature now forced to
equal the corrected temperature estimate, $\teff$.  The final adopted values of the stellar parameters for our Li-rich stars are given in
Table~\ref{tab-par}.  The error in effective temperature,
$\sigma_{\teff}=140$~K, and $\feh$, $\sigma_{\feh}=0.10$~dex, were
adopted from \citet{ruchti10,ruchti11}.  We estimated an error in
surface gravity of 0.3~dex; however, the error could be larger for the
lowest gravity stars (as is described below). 

The method of using ionization equilibrium to derive the surface
gravity is believed to be unreliable in very metal-poor stars due to
non-LTE effects \citep{ti99,k03}.  The abundance derived from the
\nion{Li}{I} line, however, is nearly independent of the adopted
surface gravity.  In \citet{ruchti10,ruchti11}, it was assumed that no
giants lie above the RGB-tip, but we do not make this assumption here
since the evolutionary stage of our giants will affect the
interpretation of our abundance results.  We therefore did not apply
any of the corrections described in \citet{ruchti11} to the surface
gravity of our Li-rich candidates.  Most of our Li-rich candidates
have $\logg>1.0$, values which were not corrected in
\citet{ruchti10,ruchti11}.  Three stars have $\logg=0.6-0.8$, for
which the correction would only be 0.1-0.2~dex, well within our
errors.  The star T9112-00430-1 has the lowest gravity, suggesting
it lies far above the RGB-tip (see Figure~\ref{fig-iso}).  The derived values of the stellar
parameters (specifically gravity) of this star are most likely
affected by large non-LTE effects.  If we were to increase its gravity
estimate by 0.5~dex, as prescribed by \citet{ruchti11}, it would lie
close to the RGB-tip.

The majority of our Li-rich candidates have $\feh<-1.8$.  The most
metal-rich star, T5496-00376-1, has $\feh=-0.63$, which is much more
metal-rich than the rest of the candidates but is at the lower end of
most previous studies.  We include it here, because of its Li-rich
nature.  Our estimates of $\teff$ and $\feh$ for M3-IV101 strongly
resemble those found by \citet{kraft99}, showing differences of only 36~K and 0.02~dex in $\teff$ and $\feh$, respectively.  Our $\logg$ value, however, is
about 0.2~dex lower than their value.  This offset is very similar to
the correction that would be made to our $\logg$ value if we were to follow
the analysis in \citet{ruchti11}.  Our $\feh$ estimate for M68-A96
also agrees within $\sim0.02$~dex with metallicity estimates for the M68 cluster
\citep{lee05}.

\subsection{Luminosity}

\begin{figure*}
\epsscale{0.75}
\plotone{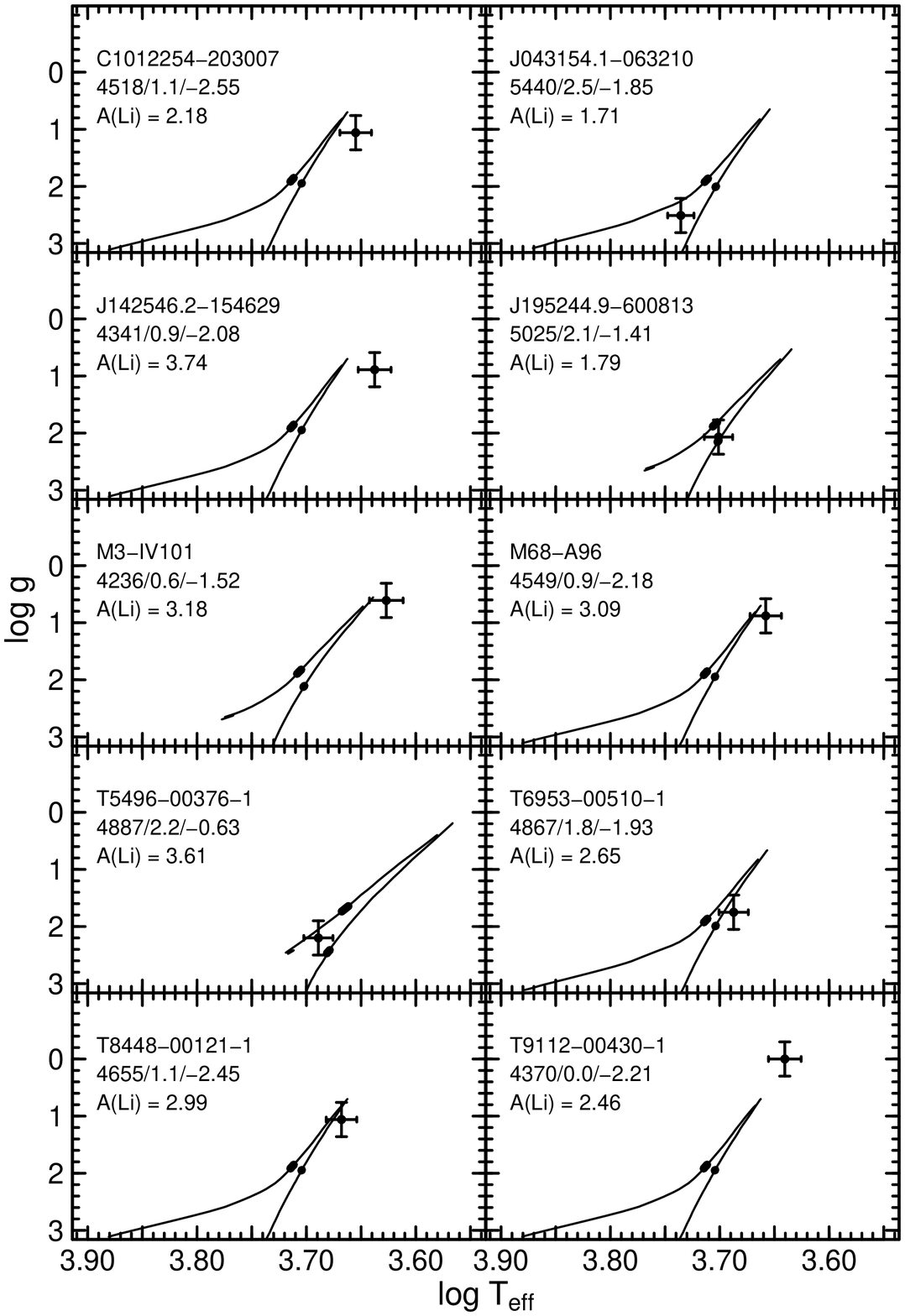}
\caption{The position of our Li-rich stars in the gravity-temperature
plane.  Also plotted are 12~Gyr Padova isochrones
\citep{marigo08,girardi10} of an appropriate metallicity to that of
each star.  The name, stellar parameters (given as $\teff/\logg/\feh$), and Li-abundance, $\ali$, are shown in each panel. The circle on the RGB and thickened region on the AGB at $\logg\sim2$ represent the luminosity bumps on the RGB and AGB, respectively.
Note that with the values estimated here, T9112-00430-1 lies well above the isochrone, which is most
likely a reflection of the inappropriateness, for this star, of the assumption of  a steady-state atmosphere in LTE.}
\label{fig-iso}
\end{figure*}

The luminosity of each star was estimated by fitting to Padova
isochrones \citep{marigo08,girardi10}.  The $Z$-metallicity of each
star was derived by combining $\feh$ and alpha-enhancement using the
transformation of \citet{salaris93}.  We assumed an alpha-enhancement
equal to [Mg/Fe] measured for each star (see \S\ref{sec-abs}).  We
then fit each star to the isochrone with the closest matching metallicity in a grid of 12~Gyr isochrones with metallicity steps of $Z=0.00002$.  It is possible that some of our stars
are younger (especially T5496-00376-1), but the error in the
luminosity due to our uncertainty in $\logg$ far outweighs this.
Figure~\ref{fig-iso} shows each star and the isochrone of the same
$Z$-metallicity in the gravity-temperature plane.

The luminosity at the point on the isochrone with the same $\logg$ value as the star being fit was chosen as the luminosity of the star, except for T9112-00430-1.  This star has a gravity above the limits of the isochrone (see Figure~\ref{fig-iso}).  We therefore adopted the luminosity at the lowest-gravity point on the isochrone.  We linearly interpolated the luminosity values versus $\logg$  for stars with $\logg$ values lying between the points on the corresponding isochrone.  We further estimated two extreme luminosity values for each star by adding and subtracting 0.3~dex to each star's $\logg$ value and then fitting again.  Errors were then estimated by taking the difference between these extreme values and the value taken from the original fit.  This resulted in a typical error in the luminosity of each star of $\sigma_{\lum}\sim0.3$.  Most stars lie between the RGB and AGB branches of the isochrones, making it difficult to label them as one or the other from inspection of Figure~\ref{fig-iso} alone.  The estimated luminosity from fitting to either branch showed no differences.  We discuss the phase of evolution of each star in more detail later (see \S\ref{sec-lisy}).  Our values for the luminosity of each star are listed in Table~\ref{tab-par}.

Given these luminosities, we followed the methodology described in \citet{ruchti11} to determine in which Galactic population each of the Li-rich giants most likely belongs.  The final population assignment for each Li-rich candidate is given in Table~\ref{tab-par}.  Note that the same population assignment was found using the stellar parameters derived in this work for the four stars (J142546.2-154629, J195244.9-600813, T5496-00376-1, and T6953-00510-1) that were also analyzed in \citet{ruchti10,ruchti11}.  All stars were assigned to the thick disk, halo, or thick/halo intermediate population \citep[see][for more details]{ruchti11}.  This implies that the Li-rich giants in our sample are old ($>10$~Gyr) and metal-poor.  Thus, they most likely have low masses, $M < 1$~\msun.

\setlength{\tabcolsep}{0.02in}
\begin{deluxetable}{rcccccc}
\tablecolumns{7}
\tabletypesize{\scriptsize}
\tablewidth{0pc}
\tablecaption{Stellar Parameter Values}
\tablehead{ \colhead{Star} & \colhead{$\teff$} & \colhead{$\logg$} & \colhead{$\feh$} & \colhead{v$_t$} & \colhead{$\lum$} & \colhead{POP\tablenotemark{\it a}} \\
                                                & \colhead{(K)}          &                                  &                               & \colhead{(\kmsec)} & & \\
              \colhead{({\it error})} & \colhead{$\pm140$} & \colhead{$\pm0.3$} & \colhead{$\pm0.10$} & \colhead{$\pm0.1$} & \colhead{$\pm0.3$} &}
                                                 
\startdata
C1012254-203007 & 4518 & 1.1 & -2.55 & 2.2 & 2.9 & 3 \\
J043154.1-063210 & 5440 & 2.5 & -1.85 & 1.3 & 1.8 & 3 \\
J142546.2-154629 & 4341 & 0.9 & -2.08 & 2.3 & 3.0 & 3 \\
J195244.9-600813 & 5025 & 2.1 & -1.41 & 1.7 & 2.0 & 2.5 \\
T5496-00376-1 & 4887 & 2.2 & -0.63 & 1.4 & 1.9 & 2 \\
T6953-00510-1 & 4867 & 1.8 & -1.93 & 1.8 & 2.3 & 2 \\
T8448-00121-1 & 4655 & 1.1 & -2.45 & 2.0 & 2.9 & 3 \\
T9112-00430-1 & 4370 & 0.0 & -2.21 & 3.1 & 3.5 & 3 \\
M3-IV101 & 4236 & 0.6 & -1.52 & 1.8 & 3.2 & -- \\
M68-A96 & 4549 & 0.9 & -2.18 & 1.8 & 3.1 & -- 
\enddata
\label{tab-par}
\tablenotetext{a}{2=thick disk, 2.5=thick/halo, 3=halo}
\end{deluxetable}

\section{Results}

\subsection{Lithium Abundances}
\label{sec-li}

The Li abundances for each star were derived assuming all the Li was
from the $^7$Li isotope.  Equivalent widths were measured for both the 6708~\AA{} and 6103~\AA{}
\nion{Li}{I} lines.  We then derived the abundance of Li from both
lines in each of LTE and non-LTE following the methods described in
\citet{lind09}.  These values can be found in Table~\ref{tab-liab}.

\begin{center}
\begin{figure*}
\epsscale{0.7}
\psfrag{Sun}{$\odot$}
\plotone{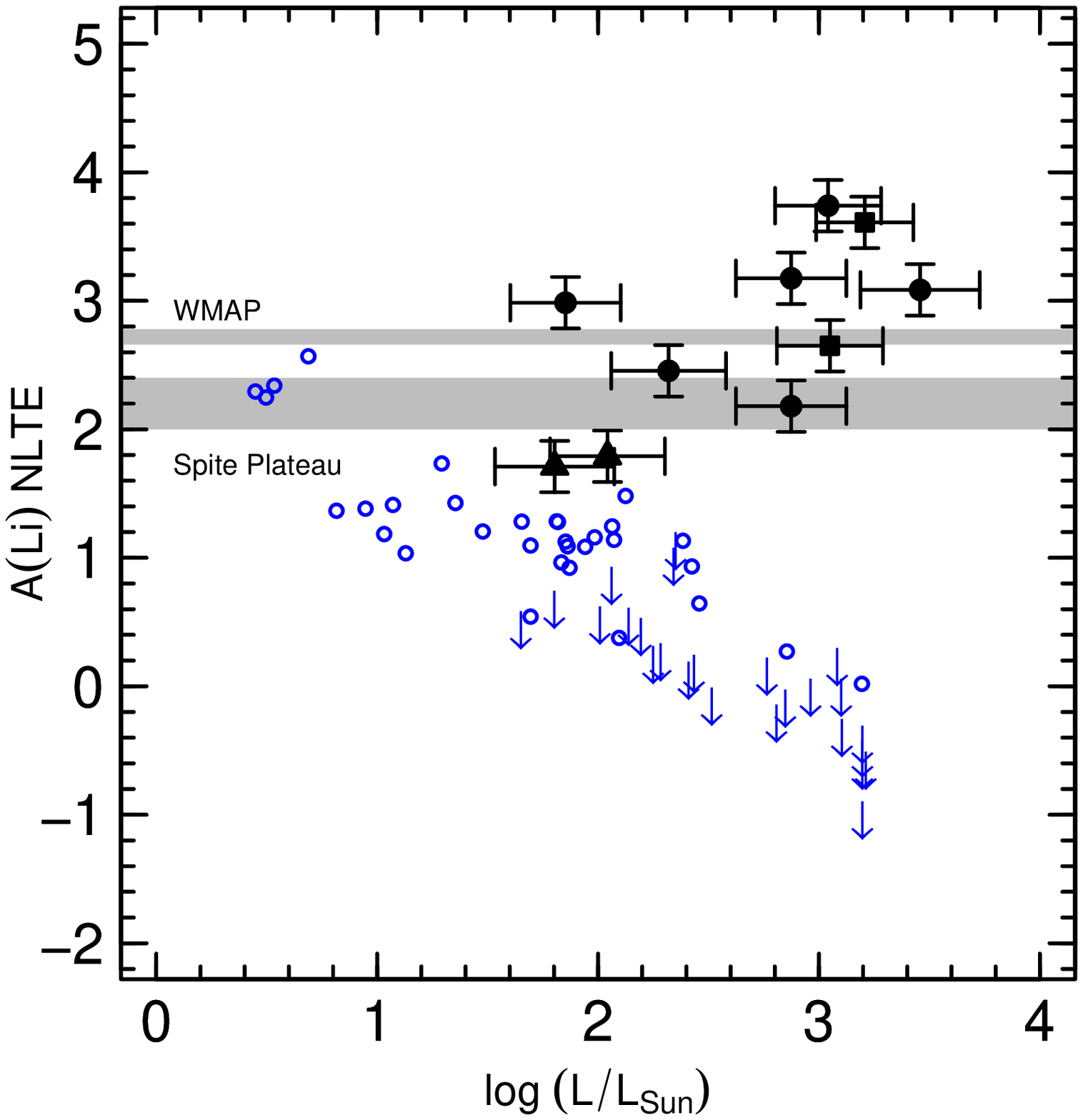}
\caption{Derived $\ali$ abundance as a function of $\lum$ for the
giants analyzed in this paper.  Stars from the RAVE-VMP sample are
plotted as blue open circles and downward pointing arrows (for upper
limits), and represent ``Li-normal" stars.  The filled circles represent 
the six Li-rich field giants discovered in this paper from the RAVE sample, and the filled squares represent
the two Li-rich globular cluster giants.  The filled triangles represent the two giants in our sample that are more Li-normal for the given 
luminosity bin.  Error bars in $\lum$ represent the difference in
luminosity when we added or subtracted 0.3~dex from our value of
$\logg$ for each star. Also shown are the range of estimates for the primordial abundance of Li given by WMAP and the Spite Plateau.  Note that several of the giants in our sample have derived Li abundances that are much greater than the primordial values.}
\label{fig-lil}
\end{figure*}
\end{center}

The derived $\ali$ values for our Li-rich 
sample are shown in Figure~\ref{fig-lil} as a function of the stars' estimated
luminousity (calculated above).  Note that, for stars that had $\ali$ estimates from both \nion{Li}{I} lines,  the value in the figure is the mean of the two values.  Further, we found a typical error of $\sim0.20$~dex from the difference between that found for the 6708~\AA{} and 6103~\AA{} lines.  Figure~\ref{fig-lil} also
includes the Li abundances we derive for a sample of 58 RAVE
very metal-poor (hereafter RAVE-VMP) stars with [Fe/H] $< -2$ from \citet{f10}.  Note that 26 measurements are only upper limits.  We followed the same analysis procedure given above for all RAVE-VMP data, and the full results for the entire sample will be published in a later paper.

Note that both J043154.1-063210 and J195244.9-600813 have luminosities and Li-abundances that place them along the trend of ``Li-normal" giants near the RGB-bump.  We therefore no longer classify them as Li-rich.  The remaining Li-rich giants have Li abundances that clearly separate them from the Li-normal giants (see Figure~\ref{fig-lil}).  As was found in previous studies, the majority of our Li-rich giants separate into two regions: the lower (near the RGB-bump) and upper RGB, separated at $\lum\sim2.4$ (see also Figure~\ref{fig-iso}).  T6953-00510-1, however, appears to lie between these two regions.

\subsection{CNO Abundances}
\label{sec-cn}

We determined the CNO abundances for our giants using MOOG, under the assumption of molecular equilibrium, since the CNO atoms can be partly bound together in molecules, especially for cooler stars \citep[see, e.g.,][]{gratton90}.

Oxygen abundances were first determined from the EWs of the \nion{O}{I}
forbidden lines at 6300~\AA{} and 6363~\AA{}, and are given in
Table~\ref{tab-liab}.  The {\it gf}-values of the lines were taken
from \citet{lambert78}, and the solar abundance, $A({\rm O})=8.69$,
was selected from \citet{asplund09}.  We corrected the \nion{O}{I}
6300~\AA{} line for the weak \nion{Ni}{I} 6300.34~\AA{} line
\citep[see][]{ap99} following the same methodology as
\citet{fulbright03}.

We next determined $\cfe$, $\nfe$, and the \ciso{} ratios for our giants by
spectral syntheses of CH and CN lines using the Plez line lists and
oscillator strengths \citep[Plez 2011, private communication; see also][]{hill02,gustafsson08} combined with
the Vienna Atomic Line Database
\citep[VALD\footnote{http://vald.astro.univie.ac.at/$\sim$vald/php/vald.php},][]{kupka00}.
We adopted dissociation energies of 3.47~eV \citep{huber79} and
7.66~eV \citep{lambert78} for CH and CN, respectively.  The CH lines
between 4320-4328~\AA{} were used to estimate the C abundance of each
star given the value of $A({\rm O})$ found above.  Given these values of $A({\rm O})$ and $A({\rm C})$, the N abundance was determined using CN lines
at the 3883~\AA{} band (and the 4216~\AA{} band, see below).  Finally,
the \ciso{} isotopic ratio was constrained by combining information from
the above lines with that derived from fits to CH and CN lines in the wavelength range 4200-4220~\AA{}.  We then iterated this procedure with the input CNO abundances equal to the previous iteration until we obtain a self-consistent solution to all three abundances.

\begin{center}
\begin{figure*}
\epsscale{0.95}
\plotone{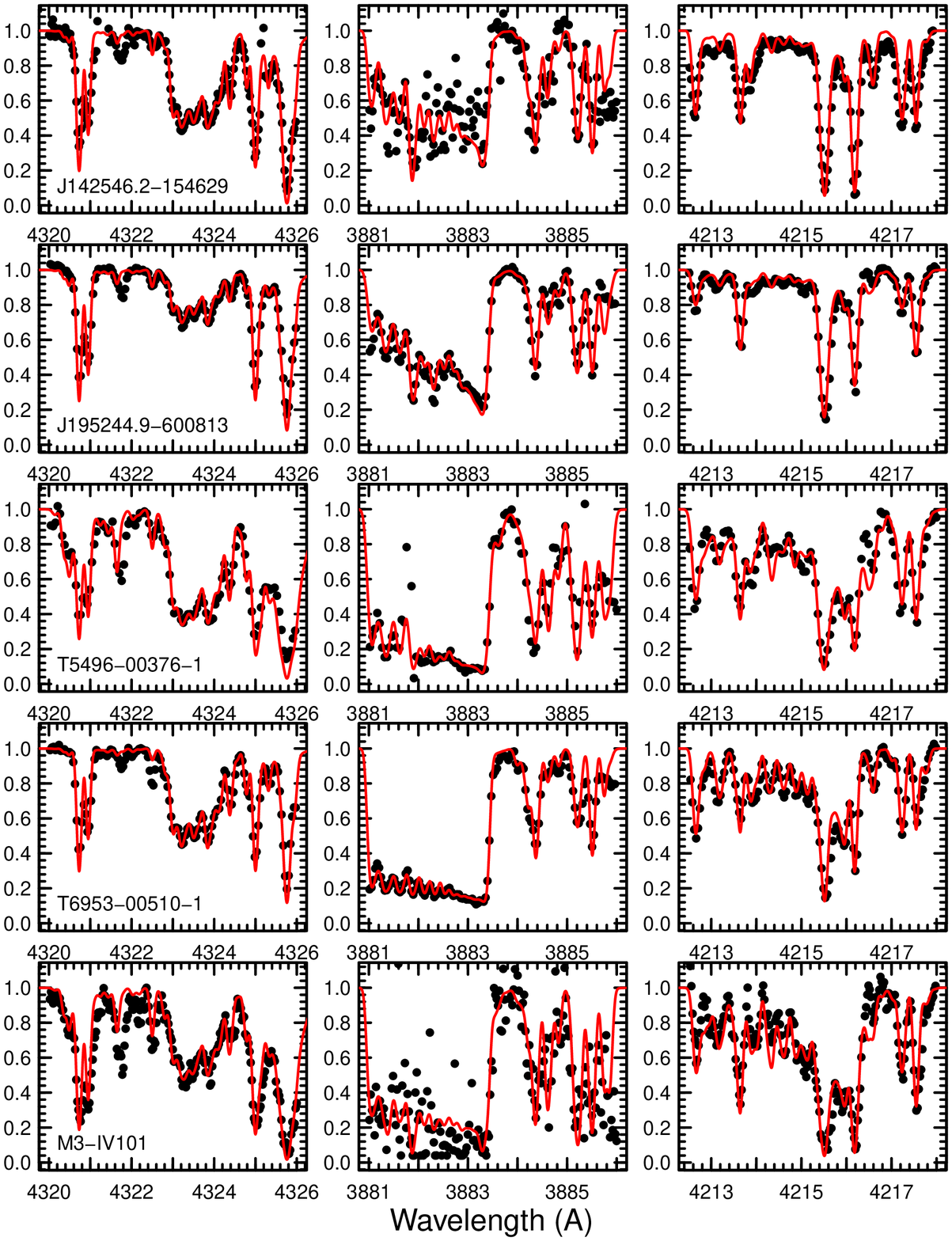}
\caption{Sample of the CN-strong stars' spectra in the region used to measure the carbon abundance (left) and the two regions used to measure the N abundance, CN-3883 (middle) and CN-4216 (right).  The solid-red line represents the synthesis with the best fit parameters given in Table~\ref{tab-liab}.}
\label{fig-cns}
\end{figure*}
\end{center}

\begin{center}
\begin{figure*}
\epsscale{0.95}
\plotone{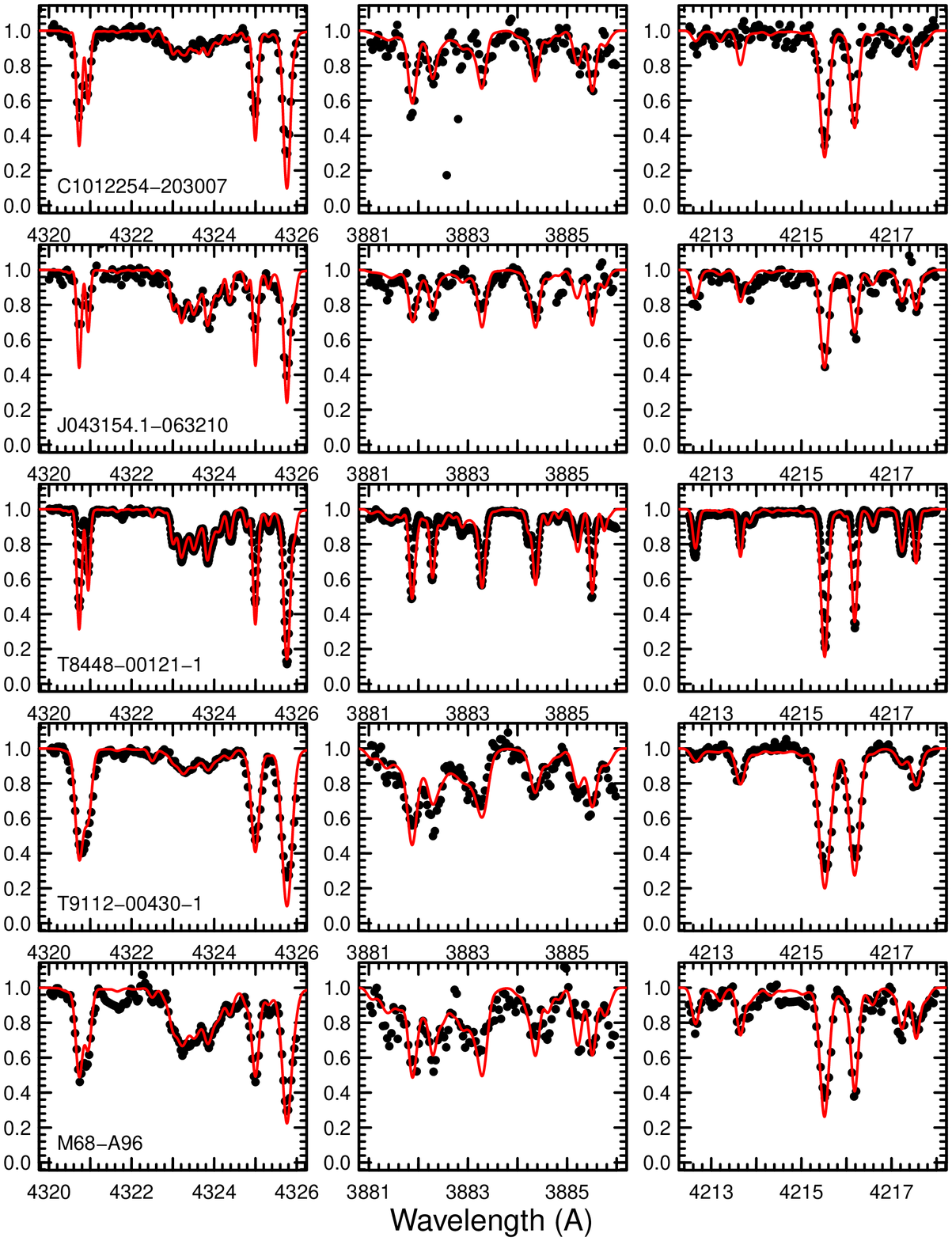}
\caption{The same as Figure~\ref{fig-cns}, but for those stars with a weak CN-3883 band.}
\label{fig-cnw}
\end{figure*}
\end{center}

During the syntheses of C and N, the Li-rich candidates fell into two classes: those with a strong CN-3883 band and those with a weak one.  The final syntheses are shown in Figures~\ref{fig-cns} and \ref{fig-cnw}, respectively.  The lines in the CN-strong stars could very well be saturated (most obvious in the M3-IV101 spectrum).  We therefore included the 4216~\AA{} CN-band in our syntheses to constrain the N abundance for these stars.  Although the CN-4216 band is very weak for the CN-weak stars, we include it in Figure~\ref{fig-cnw} for comparisons.

\begin{figure}
\epsscale{1.15}
\plotone{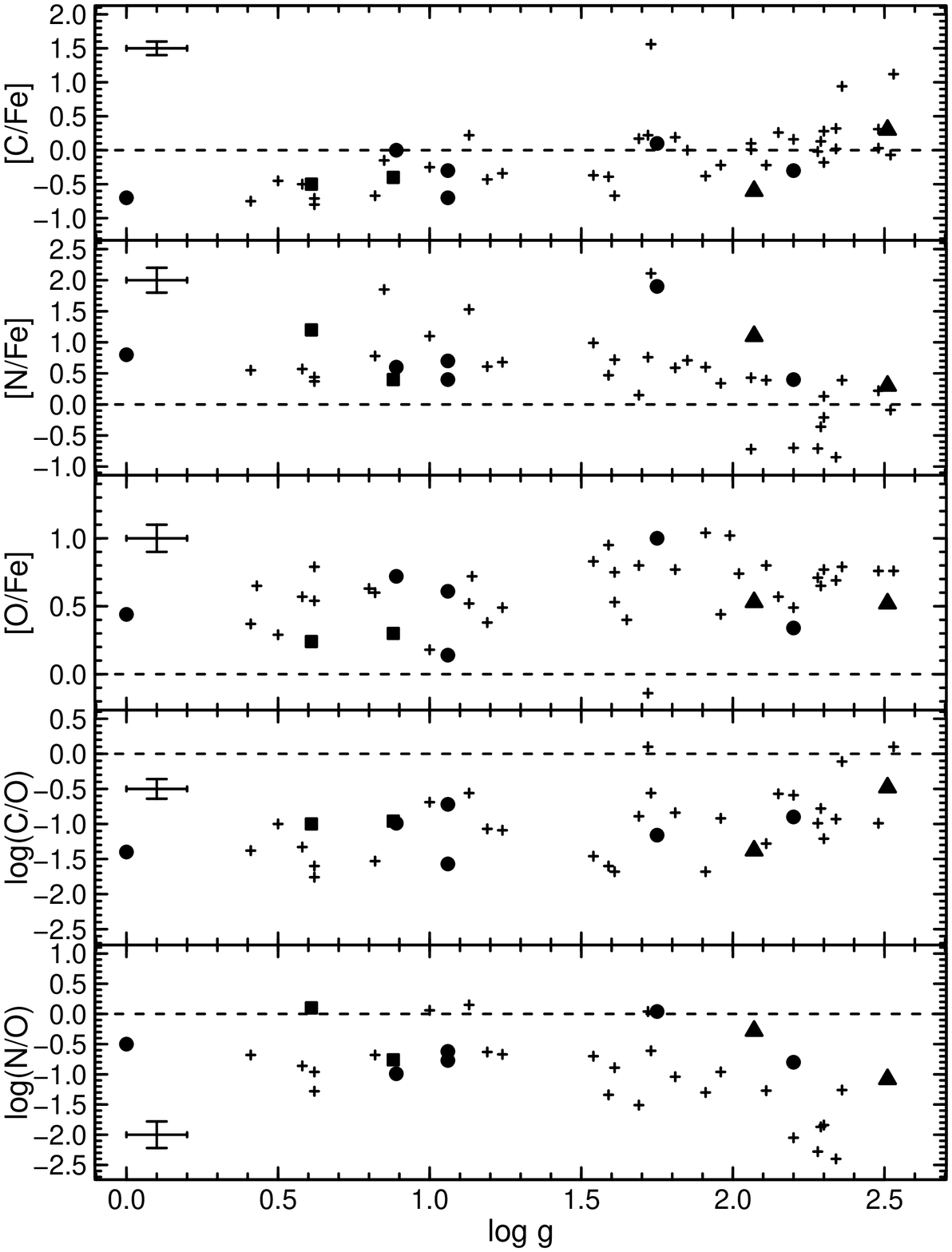}
\caption{CNO abundance ratios with Fe, as well as the log of the number ratios of  C/O and N/O, are shown versus $\logg$.  The six Li-rich field giants are represented as circles, while the two Li-normal field giants are shown as triangles.  The two Li-rich globular cluster giants are shown as squares.  The VMP comparison sample are shown as plus symbols.  Note that all of the Li-rich giants have ${\rm C/O}<1$ ($\log({\rm C/O})<0$), but two stars have ${\rm N/O}\simgt1$.  T6953-00510-1 has the highest enhancement in N and O, while J043154.1-063210 has the highest enhancement in C.}
\label{fig-cno}
\end{figure}

The resultant $\cfe$ and $\nfe$ values from the syntheses are given in
Table~\ref{tab-liab} and plotted versus $\logg$ in
Figure~\ref{fig-cno}.  Note that we adopted the solar
abundances for C and N from \citet{asplund09}.  The scatter in
individual $\cfe$ and $\nfe$ values is due to variations between
stars, not the quality of the determination (we investigate the interpretation of this scatter in a separate paper).  We estimated the internal errors to be  
$\sim0.1$~dex in $\cfe$ and $\sim0.1-0.2$~dex in
$\nfe$, depending on the quality of the spectrum and the strength of
the CN-bands.  We further varied the estimated values of the stellar parameters of each star
according to their 1-sigma errors to investigate the sensitivity of our 
syntheses to the chosen values of the stellar parameters for each star.  Our stars
consistently show errors of $\sim0.1$ and $\sim0.1-0.2$~dex in $\cfe$
and $\nfe$, respectively, and show the highest sensitivity to changes
in the effective temperature of a star.  Using the same error analysis as for C and N, we
found that our errors in $\ofe$ are about 0.10~dex.  It is important to
note, however, that the EWs for C1012252-203007, J043154.1-063210,
M68-A96, and T8448-00121-1 are less than 10~m\AA{} for both lines, and
so the error in $\ofe$ increases to $\sim0.15-0.20$~dex.  The \ciso{} isotopic ratio
was very difficult to constrain since most of our stars were fairly
deficient in carbon (see below), and so the $^{13}{\rm CH}$ features
were very weak.  The best-fit value for most stars subsequently had a
fairly flat likelihood peak.  We therefore give a range of values, as
well as the best fit value, in Table~\ref{tab-liab}.  

The CNO abundances of M3-IV101 have been analyzed previously in
the literature, which makes it useful for comparisons.  We only
compare the abundance values, since there will be offsets in the
element ratios with iron due to differences in reference solar
abundances.  \citet{kraft99} found using CH lines that $A({\rm
C})\sim6.55$ for this star, while \citet{pilachowski03} found a value
of $\sim6.31$ using CO lines.  Our value of $\cfe=-0.5$ corresponds to
$A({\rm C})=6.4$, which lies between the Kraft et al. and Pilachowski et
al. values.  The variation in the values could be attributed to
differences in atomic and molecular line data.  On the other hand, we found values of $A({\rm O})=7.41$ and $A({\rm N})=7.61$, which are only 0.01~dex and 0.06~dex higher than those found by \citet{kraft99}, respectively.  Note that
\citet{pilachowski03} did not measure N abundances in their work.
They did, however, measure the carbon isotopic ratio,
finding \ciso{}~$=11$, which is very close to our best-fit value given in
Table~\ref{tab-liab}.

The final value of $\cfe$ for the Li-rich giants ranges from $-0.7$ for C1012254-203007 and T9112-00430-1 to $+0.3$ for
J043154.1-063210, while all stars have $\nfe$ and $\ofe$ values greater than zero.  In Figure~\ref{fig-cno}, we also plot the ratios found for the RAVE-VMP comparison sample, described in \S\ref{sec-li}.  As illustrated in the figure, all Li-rich giants fall within the general trend of the RAVE-VMP stars.  The $\cfe$ and $\ofe$ ratios appear to possibly increase with gravity, while $\nfe$ ratio may decrease with gravity.  We computed Spearman's rank correlation coefficient, $r_s$, to investigate the level of correlation between all three ratios and $\logg$. We found that $\cfe$ weakly correlates with gravity, with a value of $r_s\sim+0.5$.  However, we found values of $r_s\sim+0.3$ and $-0.3$ for the $\ofe$ and $\nfe$ ratios, respectively, which implies no significant correlation.   The weak trend in $\cfe$ is a sign that the lower gravity stars have been affected by more CNO cycling and internal mixing (see also \S\ref{sec-lisy}).  The Li-rich giants at large gravity ($\logg > 1.5$) show the possibility of a slight enhancement in $\nfe$ as compared to the RAVE-VMP stars, but this is not conclusive given errors.  We will discuss these trends in more detail for the entire RAVE-VMP sample in a later paper, but the main point to take away is that the Li-rich giants and RAVE-VMP stars appear to have experienced very similar CNO-cycling.

All of the Li-rich giants have a ratio of the number of C atoms to O atoms, $\log({\rm C/O})<0$, which can be an indicator of the nature of their evolution should any of them be AGB stars (see \S\ref{sec-agb} for more details).  M3-IV101 and T6953-00510-1, however, have $\log({\rm N/O})\simgt0$.  Further, like in $\nfe$, the Li-rich giants at large $\logg$ show indications of a slight enhancement in $\log({\rm N/O})$ as compared to the RAVE-VMP stars.  Note that T6953-00510-1 is the most enhanced in $\nfe$ and [O/Fe] and J043154.1-063210 has the highest enhancement in $\cfe$ among the Li-rich giants.

\setlength{\tabcolsep}{0.04in}
\begin{deluxetable*}{rcrcccrrcccrcc}
\tabletypesize{\scriptsize}
\tablecolumns{14}
\tablewidth{0pc}
\tablecaption{Lithium and CNO Abundance Data}
\tablehead{ 
 & & & & & & & \multicolumn{3}{c}{Li-6708~\AA} & & \multicolumn{3}{c}{Li-6103~\AA} \\
 \cline{8-10} \cline{12-14} \\
\colhead{Star} & \colhead{$\feh$} & \colhead{$\cfe$\tablenotemark{a}} & \colhead{$\nfe$\tablenotemark{a}} & \colhead{$\ofe$\tablenotemark{a}} & \colhead{$\log({\rm C/O})$} & \colhead{\ciso\tablenotemark{b}} & \colhead{EW Li} & \colhead{$\ali$} & \colhead{$\ali_{\rm NLTE}$} & \colhead{} & \colhead{EW Li} & \colhead{$\ali$} & \colhead{$\ali_{\rm NLTE}$}}
\startdata
C1012254-203007 & -2.55 & -0.7 & 0.7 & 0.61 & -1.6 & $1-5$~(1) & 279.6 & 2.52 & 2.30 & & 6.3 & 1.93 & 2.06 \\
J043154.1-063210 & -1.85 & 0.3 & 0.3 & 0.52 & -0.5 & $5-15$~(12)& 30.0 & 1.69 & 1.71 & & -- & -- & -- \\
J142546.2-154629 & -2.08 & 0.0 & 0.6 & 0.72 & -1.0 & $>10$~(22) & 540.0 & 3.86 & 3.63 & & 211.6 & 3.85 & 3.85 \\
J195244.9-600813 & -1.41 & -0.6 & 1.1 & 0.53 & -1.4 & $\simgt5$~(5) & 72.7 & 1.73 & 1.79 & & -- & -- & -- \\
T5496-00376-1 & -0.63 & -0.3 & 0.4 & 0.34 & -0.9 & $5-20$~(15) & 323.7 & 3.28 & 2.94 & & 34.0 & 2.91 & 3.03 \\
T6953-00510-1 & -1.93 & 0.1 & 1.9 & 1.00 & -1.2 & $3-7$~(5) & 254.3 & 2.82 & 2.53 & & 8.8 & 2.28 & 2.38 \\
T8448-00121-1 & -2.45 & -0.3 & 0.4 & 0.14 & -0.7 & $>10$~(28) & 389.2 & 3.71 & 3.06 & & 67.5 & 3.19 & 3.29 \\
T9112-00430-1\tablenotemark{c} & -2.21 & -0.7 & 0.8 & 0.44 & -1.4 & $>5$~(10) & 485.2 & 3.15 & 3.08 & & 68.0 & 2.92 & 3.09 \\
M3-IV101 & -1.52 & -0.5 & 1.2 & 0.24 & -1.0 & $10-20$~(12) & 508.5 & 3.49 & 3.46 & & 211.6 & 3.68 & 3.76 \\
M68-A96 & -2.18 & -0.4 & 0.4 & 0.30 & -1.0 & $3-5$~(4) & 336.5 & 2.95 & 2.64 & & 23.0 & 2.52 & 2.66 \\
\enddata
\tablenotetext{a}{Solar abundances adopted from \citet{asplund09}.}
\tablenotetext{b}{Number in parentheses is the best fit value for CH and CN lines between $4200-4220$~\AA.}
\tablenotetext{c}{$\ali$ computed for $\logg=0.5$.}
\label{tab-liab}
\end{deluxetable*}

\subsection{Additional Elements}
\label{sec-abs}

We have used the line lists of \citet{fulbright00} and \citet{jj02} to measure the
abundances of other elements for these stars, including Mg, Na, and
several neutron-capture elements, for which the ratio with iron is given in Table~\ref{tab-ab}.  Hyperfine splitting effects were taken into account for the Na I D lines and the lines of Ba II and Eu II.  Solar values for all elements were again selected from \citet{asplund09}.  The neutron-capture elements can be especially important since they can indicate the possible presence of dredge-up in AGB stars.  

\begin{figure*}
\epsscale{0.8}
\plotone{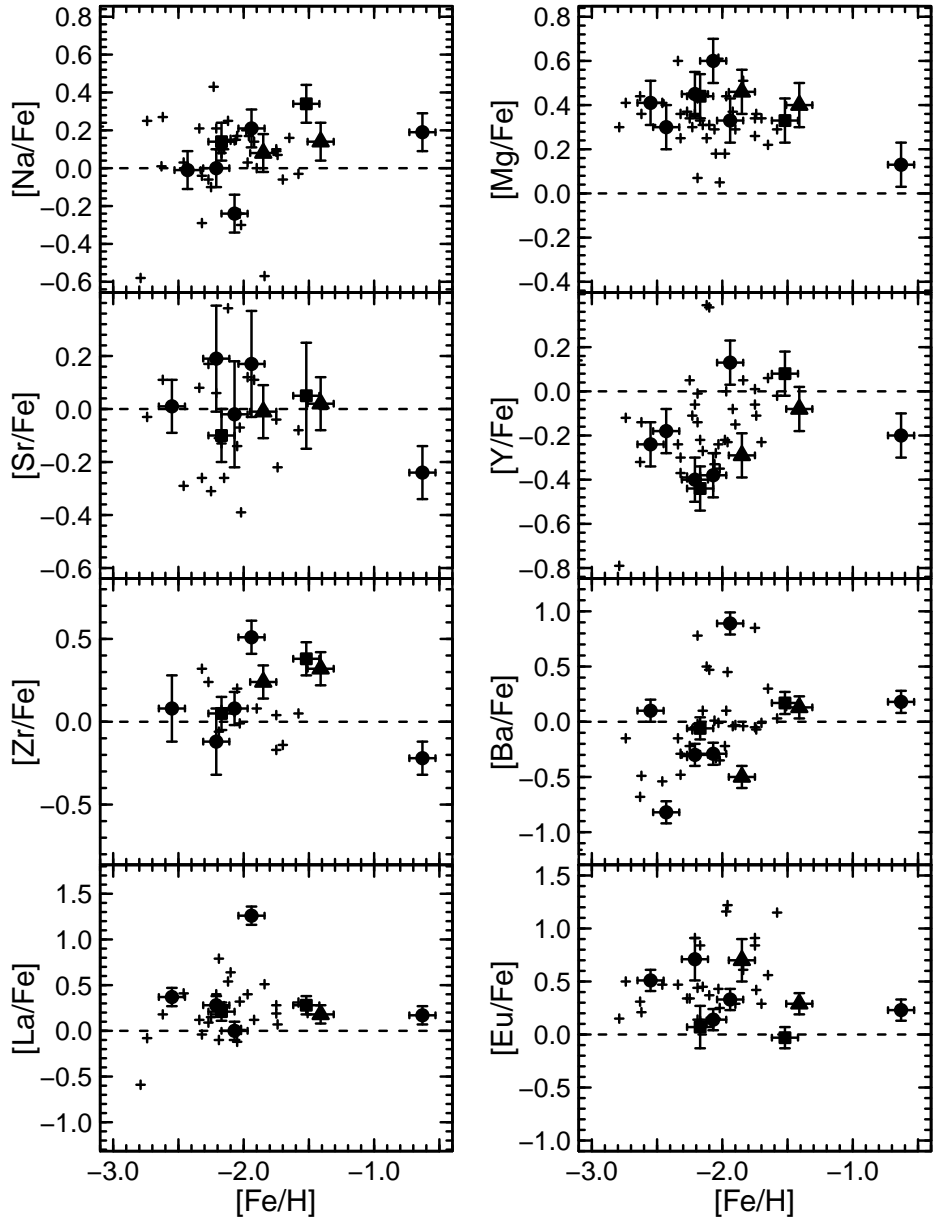}
\caption{The ratio of several elements with iron abundance vs. $\feh$.  The symbols are the same as in Figure~\ref{fig-cno}.  Pb is not shown since it could not be measured for several stars.  Note, the only Li-rich giant that shows consistent enhancement in the s-process (including Pb) is T6953-00510-1.}
\label{fig-ncfe1}
\end{figure*}

\begin{figure*}
\epsscale{0.8}
\plotone{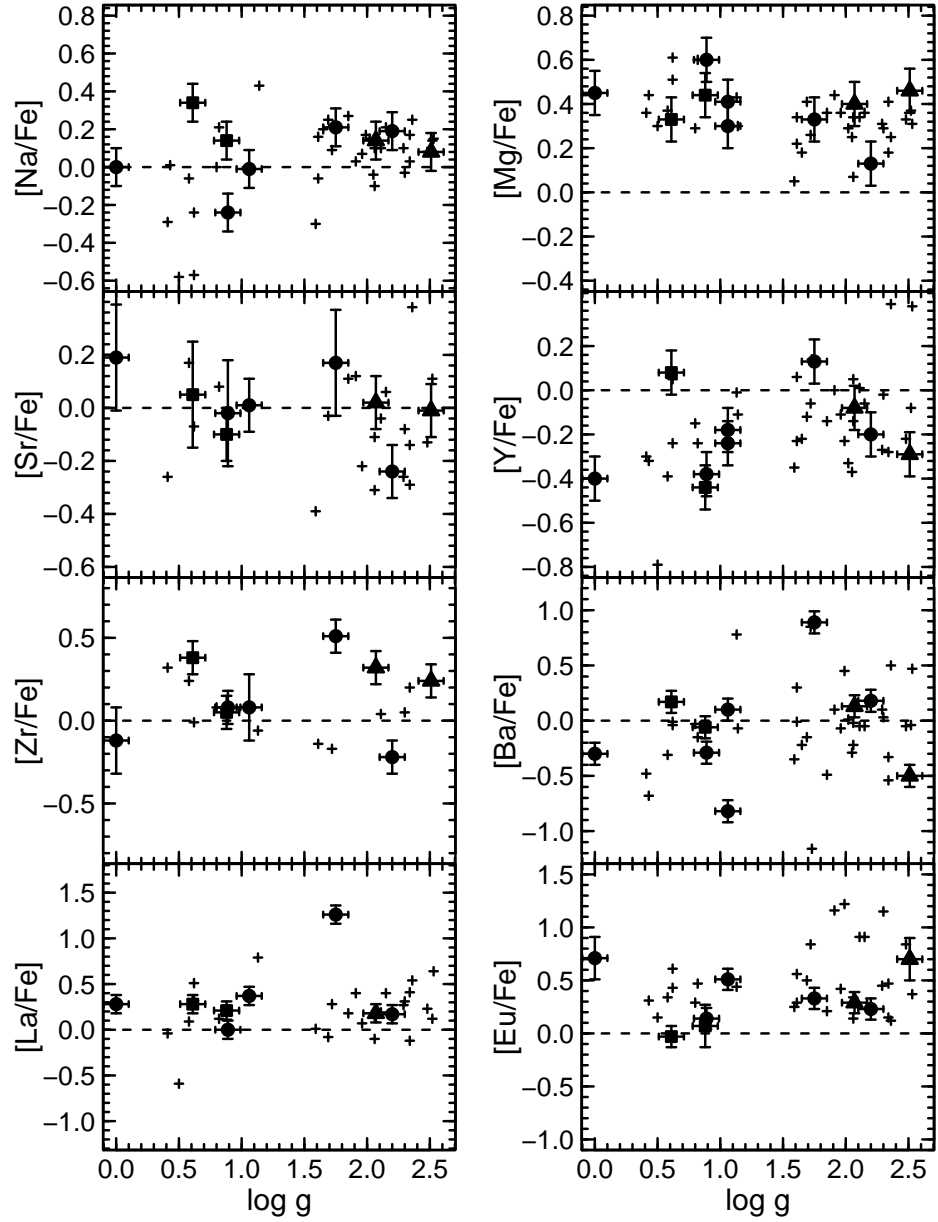}
\caption{The ratio of several elements with iron abundance vs. $\logg$.  The symbols are the same as in Figure~\ref{fig-cno}.  As in Figure~\ref{fig-ncfe1}, the Li-rich giants do not show any deviation from the general trends of the VMP comparison sample.  Note that the apparent enhancement in [Y/Fe] of M3-IV101 is a metallicity artifact.}
\label{fig-ncfe2}
\end{figure*}

\begin{figure}
\epsscale{1.15}
\plotone{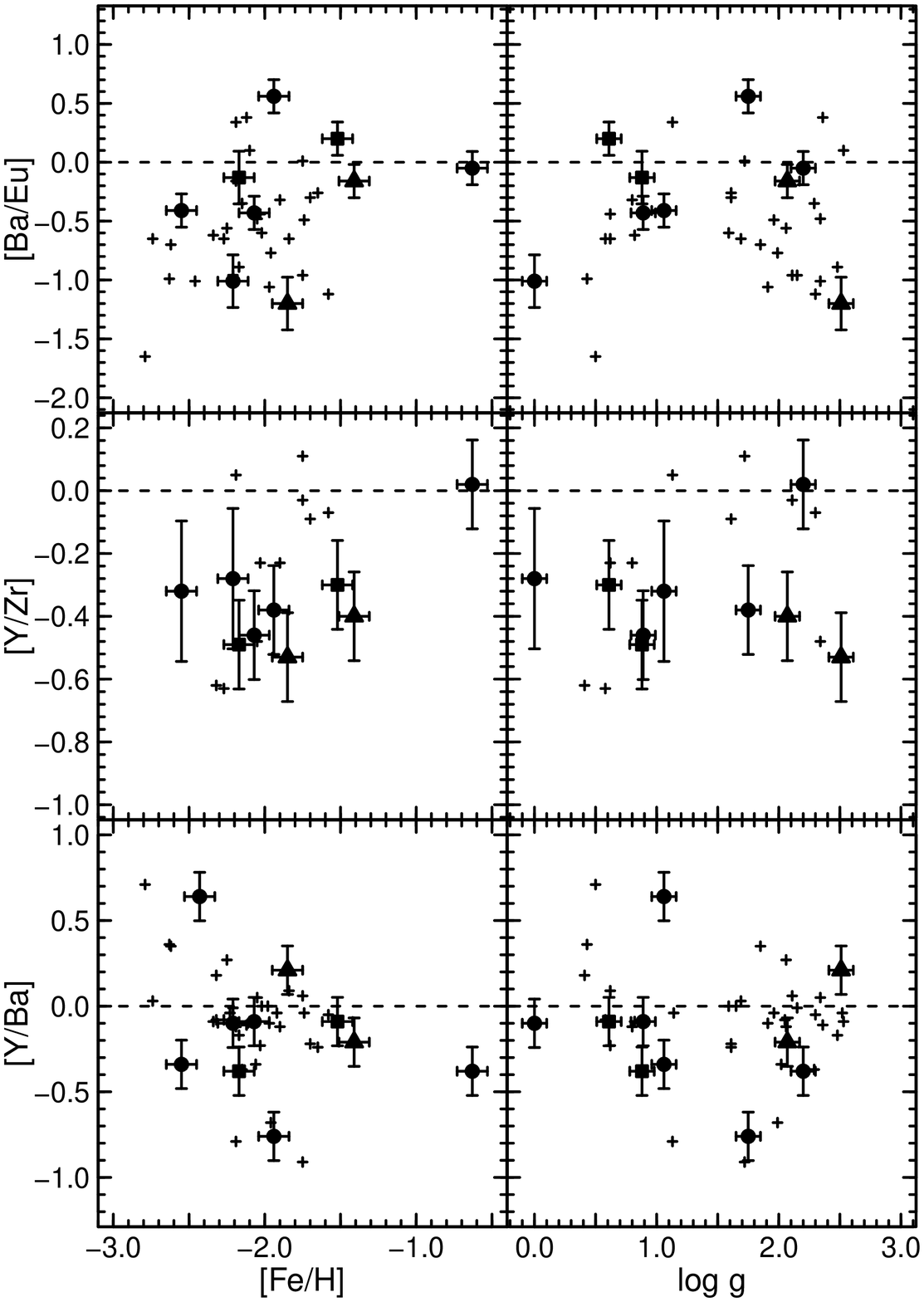}
\caption{The ratio of several neutron-capture element abundances vs. metallicity.  The symbols and lines are the same as in Figure~\ref{fig-cno}.  Note, that T6953-00510-1 shows enhancement in [Ba/Eu], suggesting (as in Figures~\ref{fig-ncfe1} and \ref{fig-ncfe2}) that it is s-process enhanced.}
\label{fig-ncrat}
\end{figure}

The derived abundance ratios for these elements are shown versus $\feh$ and $\logg$ in Figures~\ref{fig-ncfe1} and \ref{fig-ncfe2}, respectively, for our Li-rich candidates.  We further plot the abundance ratios of our Li-normal VMP giants.  Most of our stars have ratios that resemble that of the Li-normal VMP giants of like metallicity and gravity.  The Li-rich giant, T6953-00510-1, however, shows enhancement in the s-process elements (Sr, Y, Zr, Ba, La, Pb) as compared to the other stars.  Further evidence for s-process enhancement is found in the top plots of Figure~\ref{fig-ncrat}, in which T6953-00510-1 shows enhancement in [Ba/Eu], which gives the ratio of s-process to r-process in a star.  Other ratios, such as [Y/Zr], [Y/Ba], and [Ba/Eu] (see Figure~\ref{fig-ncrat}), are similar between the Li-rich giants and Li-normal VMP stars.

Since T6953-00510-1 is the only star with enhanced s-process
abundances, this enhancement is probably not connected to the
mechanism that is enriching the Li in our stars.  Further, this star
is not located near the TP-AGB evolutionary phase (see
Figure~\ref{fig-iso}) where we would expect dredge-up of s-process
enriched material. The simplest explanation is that this star is a part of a binary system in which its
atmosphere has been polluted by a higher-mass companion that has
already gone through its TP-AGB phase.  We do not detect any variation
in the radial velocity of this star (see \S\ref{sec-rv}), but this
scenario is still possible if the binary system is wide.  It is also possible that T6953-00510-1 formed
from s-process enriched material left by a star of an early generation
that had gone through its TP-AGB phase, but this scenario would require extremely incomplete mixing of the ISM prior to next generation of stars.

\citet{lee05} measured abundances for seven stars in M68, but they did not 
include M68 A-96. If we use their line list and follow their analysis methods, 
we measure abundances for this star nearly identical to their mean 
cluster values.

\begin{center}
\setlength{\tabcolsep}{0.05in}
\begin{deluxetable*}{rcrrrrrrrrrrrr}
\tabletypesize{\scriptsize}
\tablewidth{0pc}
\tablecaption{Elemental Abundance Data}
\tablehead{ \colhead{Star} & \colhead{$\feh$} &\colhead{Na\tablenotemark{a}} & \colhead{Mg\tablenotemark{a}} & \colhead{Sr\tablenotemark{a}} & \colhead{Y\tablenotemark{a}} & \colhead{Zr\tablenotemark{a}} & \colhead{Ba\tablenotemark{a}} & \colhead{La\tablenotemark{a}} & \colhead{Eu\tablenotemark{a}} & \colhead{Pb\tablenotemark{a}} & \colhead{[Ba/Eu]} & \colhead{[Y/Zr]} & \colhead{[Y/Ba]}}
\startdata
C1012254-203007 & -2.55 & -- & 0.41 & 0.01 & -0.24 & 0.08 & 0.10 & 0.37 & 0.51 & -- & -0.41 & -0.32 & -0.34 \\
J043154.1-063210 & -1.85 & 0.08 & 0.46 & -0.01 & -0.29 & 0.24 & -0.50 & -- & 0.70 & -- & -1.20 & -0.53 & 0.21 \\
J142546.2-154629 & -2.08 & -0.24 & 0.60 & -0.02 & -0.38 & 0.08 & -0.29 & 0.00 & 0.14 & -- & -0.43 & -0.46 & -0.09 \\
J195244.9-600813 & -1.41 & 0.14 & 0.40 & 0.02 & -0.08 & 0.32 & 0.13 & 0.18 & 0.29 & 0.19 & -0.16 & -0.40 & -0.21 \\
T5496-00376-1 & -0.63 & 0.19 & 0.13 & -0.24 & -0.20 & -0.22 & 0.18 & 0.17 & 0.23 & 0.22 & -0.06 & 0.02 & -0.38 \\
T6953-00510-1 & -1.93 & 0.21 & 0.33 & 0.17 & 0.13 & 0.51 & 0.89 & 1.26 & 0.33 & 1.48 & 0.56 & -0.38 & -0.76 \\
T8448-00121-1 & -2.45 & -0.01 & 0.30 & -- & -0.18 & -- & -0.82 & -- & -- & -- & -- & -- & 0.64 \\
T9112-00430-1 & -2.21 & 0.00 & 0.45 &  0.19 & -0.40 & -0.12 & -0.30 & 0.28 & 0.71 & -- & -1.01 & -0.28 & -0.10 \\
M3-IV101 & -1.52 & 0.34 & 0.33 & 0.05 & 0.08 & 0.38 & 0.17 & 0.28 & -0.03 & 0.26 & 0.20 & -0.30 & -0.09 \\
M68-A96 & -2.18 & 0.14 & 0.44 & -0.10 & -0.44 & 0.05 & -0.06 & 0.21 & 0.07 & 0.53 & -0.13 & -0.49 & -0.38 \\
\enddata
\tablenotetext{a}{Given as [X/Fe].}
\label{tab-ab}
\end{deluxetable*}
\end{center}

\subsection{Projected Rotational Velocity}
\label{sec-vsini}

Some have found that many of their metal-rich Li-rich giants had high projected rotational velocities, $\vsini$ \citep[e.g.,][]{guillout09}.  Further, \citet{drake02} suggested that the fraction of Li-rich stars can be as high as 50 percent among rapidly rotating giants.    We therefore computed $\vsini$ for the Li-rich giants in our sample to determine if any are rapid rotators.  We derived $\vsini$ following the methodology of \citet{fekel97} and \citet{hekker07}.  First, the measured stellar broadening, $X_{\rm meas}$, was estimated as the average of the full-width-at-half-maximum (FWHM) of several \nion{Fe}{I} lines near 6750~\AA{}.  The FWHM of several ThAr lines (from arc spectra taken during the night of each observation) in the same wavelength region as the \nion{Fe}{I} lines were averaged to estimate the instrumental broadening, $X_{\rm inst}$.  The intrinsic broadening can then be estimated as,
\begin{equation}
X_{\rm intr}=\sqrt{X_{\rm meas}^2-X_{\rm inst}^2}.
\end{equation} 
We then determined the total broadening, $X_{\rm tot}$, given our value of $X_{\rm intr}$ for each star, using the second-order polynomial fit to $X_{\rm intr}$ vs. $X_{\rm tot}$ from \citet{hekker07},
\begin{equation}
X_{\rm intr}=0.10963 + 0.002758X_{\rm tot} + 0.001278X_{\rm tot}^2.
\end{equation}

The projected rotational velocity, $\vsini$, of a given star can then be computed as $\sqrt{X_{\rm tot}^2-v_{\rm m}^2}$, where $v_{\rm m}$ is the macro-turbulent velocity of that star.  We adopted the relations between $v_{\rm m}$ and $\teff$ in \citet{hekker07} for different luminosity classes to estimate $v_{\rm m}$ for each Li-rich giant.  We assumed that the Li-rich giants in our sample near the RGB-tip were class~II and those near the RGB-bump were class~III.  Note that those stars with estimated values of $v_{\rm m}$ greater than the total broadening are assumed to have a non-measurable rotational velocity.

The majority of our Li-rich giants showed no measurable rotational velocity.  We found non-negligible values of $\sim3$~\kmsec~for T5496-00376-1 and $\sim4$~\kmsec~for T8448-00121-1, but these values are only slightly larger than the expected projected rotation of low-mass giants, $\vsini\leq2$~\kmsec~\citep{demedeiros96}. C1012254-203007 and T9112-00430-1, however, have $\vsini=8-10$~\kmsec.  This implies that they very well could be rapidly rotating.  

What would cause these stars to rapidly rotate?  It is possible that extra angular momentum, dredged-up as mass is redistributed during increased convection, could induce increased rotation in these stars \citep{fekel93}.  Another possibility is that these stars have accreted material from a planet or companion star, which would also contribute extra angular momentum to the stars.  If indeed the two rapidly rotating stars are in a binary system, signatures may be present in their spectra, which is discussed below.

\subsection{Radial Velocity Variations}
\label{sec-rv}

None of the echelle data shows obvious line profile
asymmetries or other spectral signs of having a bright secondary star
in the system, but the light from a low-luminosity main sequence or
white dwarf companion would be swamped by the light from the bright
giant star in the optical spectrum.

For most of our stars, we only have two radial velocity measurements:  the
initial RAVE DR3 \citep{siebert11} observation (with a systematic uncertainty of about 3 \kmsec) and
the echelle observation discussed here.  RAVE observed two of the stars
twice, and we have three echelle observations of C1012254-203007.
All together, we have 21 radial velocity measurements of the eight RAVE Li-rich
stars.  The mean difference (RAVE$-$echelle) in the heliocentric radial velocities is 
$0.8\pm1.8$~\kmsec, which lies within the RAVE RV measurement error.

Both T9112-00430-1 and T6953-00510-1 shared the largest difference of 3.8~\kmsec{} and 3.4~\kmsec{}, respectively, between two observations of the same star.  These differences, however, are only $\sim0.5-1.0$~\kmsec{} larger than the systematic uncertainty of the RAVE radial velocities. All other repeat observations have a velocity difference of less than twice 
the internal uncertainties.

\subsection{Mass Loss}
\label{sec-mass}

We further searched the spectra of our candidate Li-rich giants for signs of possible mass-loss, inspecting the strong photospheric lines in each spectrum, such as H$\alpha$, the Na~D lines, and \nion{Ca}{II} H and K lines \citep{balachandran00,drake02}.  The only two stars in our sample to show possible evidence of mass-loss in our sample are T9112-00430-1 and J142546.2-154629.  In both cases, we identified emission in the wings of H$\alpha$.  However, the emission was less pronounced in the spectrum of J142546.2-154629.  Both stars showed no other signs (in the Na~D or \nion{Ca}{II} H and K lines) of mass-loss in their spectra.  The emission features could be an indicator that these stars are evolving on the AGB (see \S\ref{sec-agb}).  Further, this mass-loss could be a possible trigger for Li-production in these stars \citep{dla96,dla00}, however, an investigation of infrared excess is needed to confirm this.

\section{Lithium Production from RGB to AGB}
\label{sec-lisy}

The stage of evolution (AGB or RGB) in which each star belongs is
an important aspect for understanding the mechanisms that underlie Li
production.  We therefore investigated the abundance patterns in
each star in an attempt to find indicators for their evolutionary
status, as well as Li-production mechanism.

\subsection{Li-enrichment on the RGB}
\label{sec-rgb}

Early on in a low-mass star's climb up the RGB, it undergoes the first
dredge-up on the RGB after its outer convective envelope
reaches the shell-burning regions.  Standard theories of low-mass stellar evolution predict that
the \ciso{} ratio drops from over 60 to about 40 around the point of
the first dredge-up \citep{g00}.  The first dredge-up mixes fresh H on the surface layers to the interior of the star, which causes a strong molecular weight
discontinuity at the deepest extent of the convective envelope.  It
has been suggested that this molecular weight discontinuity, combined
with rotation, induces extra mixing, such as thermohaline mixing,
that triggers Li-enrichment at the RGB-bump \citep{cb00,ulrich72,kippenhahn80,charbonnel07,eggleton08}.  Further,
\citet{charbonnel07} found that for low-mass ($M<0.9M_{\odot}$),
metal-poor ($\feh<-0.5$) stars, carbon and \ciso{} are also reduced while the
nitrogen abundance increases.  These trends predicted by thermohaline mixing were shown by \citet{angelou11} to agree with observational data in the study of the CNO abundances in M3 stars.

The two Li-normal giants (J043154.1-063210 and J195244.9-600813) are
most likely evolving along the RGB.  J043154.1-063210 lies at a higher
gravity than the RGB luminosity bump (see Figure~\ref{fig-iso}).
Should it be on the RGB, it is most likely going through its first dredge-up.
J195244.9-600813, on the other hand, appears to lie on the RGB-bump.  It
is possible that both stars are at the beginning of the RGB-bump mixing
phase.  This would also explain the intermediate (best-fit) value of
the \ciso{} ratio ($\sim12$) of J043154.1-063210 since the mixing has
not yet reduced the ratio \citep[][and references therein]{charbonnel98,gratton04}.

The picture is more complicated, however, for our Li-rich giants.  The majority of these stars have best-fit \ciso{} values greater than 10.  This is unexpected beyond the RGB-bump, given the above scenario, especially once a star has reached the RGB-tip.  One possibility is that our best-fit \ciso{} values are overestimated due to a flat likelihood peak.  For example,  T5496-00376-1, the most metal-rich ($\feh=-0.63$) giant in our sample, is the only Li-rich giant with stellar parameter values consistent with the RGB-bump.  The best fit carbon isotopic ratio for this star, however, is a bit larger ($\sim15$) than expected if Li-enrichment took place as described above.  Our lower-limit to \ciso{}, however, is 5, which is much closer to that predicted by Li-enrichment at the RGB-bump.  Alternatively, this star may not have finished this mixing episode, which would imply that the \ciso{} ratio is still decreasing, and that its Li abundance is still increasing.  However, this solution would not explain those stars near the RGB-tip that would have finished the Li-enrichment phase at the RGB-bump.

The Li-enrichment obtained at the RGB-bump is also expected to decrease as a star evolves beyond the RGB-bump due to normal Li-dilution.  Indeed, the Li-rich phase at the RGB-bump is quite short.  According to \citet{dh04}, the phase should last no longer than a few million years.  This implies that the Li should be (at least partially) depleted by the time a star reaches the RGB-tip.  According to the H-R diagram, T6953-00510-1 has a gravity and temperature consistent with it being on the RGB, but it has most likely evolved past the RGB-bump.  Its Li-abundance is less than T5496-00376-1, as well as those found for solar metallicity stars \citep[cf.,][]{kumar11}.  It could therefore have ended the extra mixing phase at the RGB-bump and its Li-abundance is now being depleted.  Also note that it has a low carbon isotopic ratio, \ciso{}~$\sim5$, suggesting that it has gone through thermohaline mixing.

It should be noted that this star is the only one in our sample that shows
consistent s-process enhancement (as discussed in \S\ref{sec-abs}).
We also argue that it may be ``CNO-increased", in that its CNO
abundances are somewhat enhanced in comparison to the other Li-rich
stars in our sample.  This star is not evolved enough to have produced
(and dredged-up) the s-process and CNO abundances in its atmosphere.
Most likely these enhancements are due to pre-enrichment at birth or mass transfer in a
long-period binary.  Although we found no conclusive disparities in the radial
velocity of T6953-00510-1 (see \S\ref{sec-rv}), radial velocity
variation from a long-period binary would be virtually undetectable
without a dedicated radial velocity study.

The high-luminosity Li-rich giants in our sample, however, have Li-abundances equal to or greater than the stars in our sample near the RGB-bump (as well as that of solar-metallicity stars in previous samples), contrary to that predicted for Li-enrichment at the RGB-bump  \citep[e.g.,][]{charbonnel05,kumar11}.  A solution to this discrepancy is that these stars have undergone Li-enrichment via extra-deep mixing combined with CBP.  \citet{sackmann99} show that CBP can take place anywhere along the RGB, while the amount of Li-enrichment highly depends on the rate of mixing in the star.  They further predict that the maximum Li abundance attained by a star can occur before significant amounts of $^{13}$C lower the \ciso{} ratio.  Thus, Li-rich giants could have values of \ciso{} ranging from $\sim4$ up to $\sim30$, which is consistent with the values of \ciso{} found for our Li-rich giants.  Note also, that the projected rotational velocities ($\vsini=8-10$~\kmsec) of C1012254-203007 and T9112-00430-1 may be large enough to induce the high extra-mixing rates needed to achieve Li-enhancement via CBP. 

Another possibility is the so-called ``Li-flash'' of
\citet{p01}.  This model also predicts that the Li-rich phase begins
at the RGB luminosity function bump when $^7$Be, transported from the
interior, decays into $^7$Li and burns in a Li-burning shell.  This
extra energy rapidly increases the star's luminosity and forces extra
mixing, including the transport of more $^7$Be (which decays into
$^7$Li) to the surface.  This mixing initially does not change the
surface carbon isotope ratios, but the enhanced convection eventually
reaches the depth where $^{12}$C is converted to $^{13}$C and the
temperature is high enough to burn the freshly-minted Li (${\rm
^7Li(p,\alpha)\alpha}$).  Both the surface Li-abundance and \ciso{}
ratio are then lowered.

\citet{pilachowski03} analyzed the previously discovered Li-rich giant M3 IV-101
and found that the intermediate value of the star's \ciso{} ratio (of which we found a similar value) and its  
high luminosity were consistent with the Li-flash model.  Our analysis shows that several of the Li-rich giants are much brighter than the RGB-bump and indeed that these stars have an estimated luminosity that is about 15 times the RGB-bump luminosity.  The Li-flash model of \citet{p01} predicts a factor of $\sim 5.5$ increase in luminosity, but based on an assumed 1.5~M$_\odot$ solar-metallicity star, rather than the lower-mass VMP stars studied here.  Further, recall that \citet{dh04} showed that canonical extra mixing cannot be responsible for the Li-flash.  However, if this scenario were possible in low-mass stars, then it could explain the luminous Li-rich giants in our sample, but cannot provide a solution for the less-luminous giants.

\subsection{Li-enrichment on the Horizontal Branch}
\label{sec-hb}

\citet{kumar11} found that their stars close to the RGB-bump were more likely associated with the theoretical position of the red clump, in which metal-rich (or young) stars have begun their core He-burning.  Since it is highly unlikely that the Li-rich phase at the RGB-bump would
last until the red clump, they postulated that the Cameron-Fowler mechanism
may also play a role during the He-core flash for stars with
$M<2.25M_{\odot}$.

This relies on the fact that enough $^3$He has
survived mixing along the RGB (e.g., thermohaline mixing) for the
Cameron-Fowler mechanism to take effect.  They suggest that stars experiencing this mechanism would survive as Li-rich giants for about 1\% of their horizontal branch lifetime, which corresponds to a few Myr.  Our stellar parameter values for T5496-00376-1 show that it is also consistent with the horizontal branch (see Figure~\ref{fig-iso}).  It is therefore possible that  T5496-00376-1 could
actually be a horizontal branch (core He-burning) star that has undergone Li-production at the He-core
flash.

\subsection{Li-enrichment on the AGB}
\label{sec-agb} 

The high-luminosity stars in our sample also appear to be consistent
with the early TP-AGB phase of evolution.  T9112-00430-1, for example,
is most likely already evolving on the AGB.  Its position far
above the RGB-tip (see Figure~\ref{fig-iso}) suggests that the derived values of the stellar parameters are being affected by strong departures from
hydrostatic equilibrium.  More importantly, T9112-00430-1 has been
found to be a variable star according to the All Sky Automated Survey
(ASAS) Catalog of Variable Stars \citep{pojmanski02}. This is further
corroborated by the presence of emission in the wings of H${\alpha}$
in its spectrum (see \S\ref{sec-mass}), which suggests that it could be surrounded by a circumstellar envelope.

Hot bottom burning is not expected for our stars.  They are metal-poor, which implies they could be old, and so have masses low enough ($M\simlt1~M_{\odot}$) that the convective envelope and H-burning shell are not connected.  However, given that enough $^3$He remains after the stars ascent of the RGB, CBP (as described in \S\ref{sec-intro}) can occur on the AGB.  Extra deep mixing mechanisms, similar to that of an RGB
star (e.g., thermohaline mixing), in the radiative layer of a low-mass
AGB star can connect the H-burning shell with the convective envelope,
which will then drive the production of $^7$Li by the Cameron-Fowler
mechanism.  It is possible, however, that the Li-enrichment from CBP on the AGB might be less than that on the RGB since the amount of $^3$He is expected to be depleted from the RGB.  As noted in \S\ref{sec-rgb}, C1012254-203007 and T9112-00430-1 have projected rotation velocities large enough  that efficient mixing speeds could also be achieved to produce Li-enhancement via CBP on the AGB.

During the TP-AGB phase, third dredge-up episodes enrich
the envelope with s-process elements that are synthesized during the
period between thermal pulses, when $^{13}$C in the star's
interior is burned and supplies neutrons \citep{straniero95}.
Our high-luminosity Li-rich giants show, however, no signs of
s-process enrichment.  This implies that if they are AGB stars, they
have not gone through enough TDU episodes to drive up the s-process.
Further, \citet{girardi10} found that a low-mass ($M_{\rm
ini}\simlt1.0M_{\odot}$), low-metallicity AGB star would remain O-rich
($\log({\rm C/O})<0$) for its entire TP-AGB evolution, which is
consistent with the CNO abundances found for the high-luminosity
Li-rich giants in our sample.

We conclude that those stars in our sample that are evolving on the AGB (e.g., T9112-00430-1) should be (early) TP-AGB objects of low mass where very efficient CBP can occur, while the core is not massive enough to drive the third dredge-up.

\subsection{External Interactions}

\citet{dh04} suggest that the influence of a nearby companion (giant
planet or star) could induce the extra mixing necessary to bring
$^7$Be up to the cooler surface where Li is produced, by the reaction ${\rm
^7Be({\it e}^-,\nu)^7Li}$, and does not suffer rapid destruction from
proton captures.  Indeed, Li-rich dwarf stars that show possible lithium pollution from a companion star have been discovered in globular clusters \citep[see, for example,][]{koch11,monaco11b}.  In particular, \citet{koch11} found a Li-rich star that showed no enrichment in s-process to be in a binary system.  They suggested that the Li-enrichment arose from mass transfer from a companion giant star which had undergone CBP.

The high projected rotational velocity found for C1012254-203007 and T9112-00430-1 could have been produced by the transfer of angular momentum from a binary companion. While we did not have the observations to detect
extrasolar planets around our stars, we checked the \citet{schneider11} extrasolar planet database but did not find coincidences.  We further looked for radial velocity
variations that might indicate a low-luminosity stellar-mass
companion.  The lack of large radial velocity variations over the
whole sample strongly suggests that a close stellar-mass companion is
not required for the Li-rich phase to occur.  The abundances of T6953-00510-1, however, suggest that it may have been enriched through mass transfer by an evolved stellar companion in a wide binary.  For this reason, recent Li-enrichment (so that the Li has not yet burned away) from an evolved companion cannot be ruled out for this star.  

\section{Conclusion}

This work presents the largest sample of metal-poor Li-rich giants to date.  We have discovered five new metal-poor ($\feh\simlt-1.9$) Li-rich giants, and one Li-rich giant at a larger metallicity of $\sim-0.6$~dex in the RAVE survey.  These stars were found in a total sample of $\sim700$ stars.  This is consistent with the previous finding that about 1\% of all giants are Li-rich, which suggests that the frequency of Li-rich giants is independent of metallicity.  

We have further analyzed the newly-discovered (in this work) Li-rich member of the globular cluster M68, and confirmed the large Li abundance of M3-IV101, whose Li-rich nature was discovered by \citet{kraft99}.  The Li-rich giant in M68 adds to the very small number of giants in globular clusters identified to have Li-enrichment.  Indeed, \citet{pilachowski00} did not find any Li-rich giants in a sample of over 200 giants selected from several globular clusters.  Further, all Li-rich giants identified in globular clusters (including the giant in M68) have been found to be evolving near the RGB-tip.

We performed a detailed abundance analysis of all stars and found that, aside from Li, the majority of the Li-rich giants in our sample have abundance trends that resemble that of the RAVE-VMP comparison sample.  This is consistent with, and extends, the findings of \citet{castilho00} for solar-metallicity Li-rich giants.  The Li-rich giants in our sample are relatively carbon-poor and nitrogen-rich, with normal oxygen abundances found for metal-poor halo giants.  Only one star in our sample, namely T6953-00510-1, shows enhancements in C, N, and the s-process elements.  We attribute these enhancements to either pre-enrichment or binary pollution, and conclude that such enhancements are not connected to the mechanisms that produce Li in our full sample.  Although we found no large radial velocity variations for the stars in our sample, a radial velocity study, over a longer time-period, would
be useful to determine if any of our giants belong to a long-period binary.  We further computed the projected rotational velocities of our stars, finding that only two stars, C1012254-203007 and T9112-00430-1, are rapidly rotating with $\vsini>8$~\kmsec.  

An important finding is that the high-luminosity giants can have Li abundances equal to those found for giants near the RGB-bump, which has also been found for stars in other environments and higher metallicities \citep{dominguez04, kraft99, monaco08, monaco11}.    If the stars at the RGB-bump and RGB-tip were enriched in Li by the same process, this would argue against a single Li-enrichment phase at the RGB-bump.  In this case, the RGB-tip giants should have less Li abundances than those at the RGB-bump, contrary to our results.  Instead, the most likely scenario is that the Li-rich giants have undergone Li-enrichment via cool bottom processing.  This process is also in agreement with our best-fit \ciso{} values for the Li-rich giants.  Further, the metallicity and Galactic population membership (e.g., thick disk and halo) of the Li-rich giants in our sample are consistent with old ages and low masses ($<1$~\msun).  Our identification of luminous Li-rich giants evolving above the RGB-bump therefore contrasts with the \citet{cb00} suggestion that Li-rich giants should be found above the RGB-bump only in intermediate-mass stars.  

It is possible that some of the luminous giants in our sample are AGB stars.  In those cases, CBP is still feasible as long as some $^3$He has survived after the RGB.  Hot bottom burning is not possible for our stars, as this process requires a much larger stellar mass.  We also cannot rule out the possibility that T5496-00376-1 is a horizontal branch star that was enriched in Li during its He-flash phase, but this has not been modeled.  Future infrared observations of these stars would be beneficial for measuring any infrared-excess that would imply that they are evolving on the AGB.  In addition, asteroseismology observations obtained with the Kepler spacecraft could also be used to distinguish between RGB and HB stars \citep[e.g.][]{bedding11}. 

It is clear from this study that metal-poor Li-rich giants are crucial for constraining the models of Li-production in giants.  The discovery and analysis of more metal-poor Li-rich giants will allow for more robust statistics and enhance our understanding of these rare and important objects. 

\acknowledgements
We would like to thank the staff members of Siding Spring Observatory, La Silla Observatory, 
Apache Point Observatory, and Las Campanas Observatory for their assistance 
in making these observations possible.  GRR wishes to thank Karin Lind, Luca Sbordone, and Peter Cottrell for their useful discussions and suggestions.  GRR further thanks Bertrand Plez for kindly providing his molecular line lists and Luca Casagrande for providing his photometric temperatures.  GRR and RFGW acknowledge support from the NSF of the USA (AST-0908326).  JPF wishes to thank Jennifer Johnson
and Howard Bond for their useful suggestions, and acknowledges support 
through grants from the W. M. Keck Foundation and the Gordon and Betty Moore
Foundation, to establish a program of data-intensive science at the Johns
Hopkins University.  EKG was partially supported by the Sonderforschungsbereich ``The Milky Way System'' (SFB 881, subproject A5) of the German Research Foundation (DFG).  This publication makes use of data products of the
Two Micron All Sky Survey, which is a joint project of the University of
Massachusetts and IPAC/Caltech, funded by NASA and the NSF.  This research
has also made use of the Vizie-R databases, operated at CDS, Strasbourg, France.
Funding for RAVE has been provided by: the Australian Astronomical
Observatory; the Leibniz-Institut f\"ur Astrophysik Potsdam (AIP); the Australian National
University; the Australian Research Council; the French National Research
Agency; the German Research Foundation; the European Research Council 
(ERC-StG 240271 Galactica); the Istituto Nazionale di
Astrofisica at Padova; The Johns Hopkins University; the National Science
Foundation of the USA (AST-0908326); the W. M. Keck foundation; the
Macquarie University; the Netherlands Research School for Astronomy; the
Natural Sciences and Engineering Research Council of Canada; the Slovenian
Research Agency; the Swiss National Science Foundation; the Science \&
Technology Facilities Council of the UK; Opticon; Strasbourg Observatory;
and the Universities of Groningen, Heidelberg and Sydney. The RAVE web site is at http://www.rave-survey.org.

{\it
\facility
Facilities: \facility{ARC (echelle spectrograph)}, \facility{AAT (UCLES)}, \facility{Magellan:Clay (MIKE)}, \facility{Max (FEROS)}, \facility{UKST (6dF spectrograph)}
}

\end{document}